\documentclass[twocolumn]{aastex631}

\newcommand{\hst}{{\it HST}}
\newcommand{\masyr}{\,\mathrm{mas}\,\mathrm{yr}^{-1}}
\newcommand{\gaiahub}{\textsc{GaiaHub}}

\submitjournal{ApJ}

\shorttitle{Proper Motions and Orbits of Distant Local Group Dwarf Galaxies}
\shortauthors{Bennet et al.}

\graphicspath{{./}{}}

\begin{document}

\title{Proper Motions and Orbits of Distant Local Group Dwarf Galaxies\\ from a combination of \textit{Gaia} and Hubble Data}

\correspondingauthor{Paul Bennet}
\email{pbennet@stsci.edu}

\author[0000-0001-8354-7279]{Paul Bennet}
\affiliation{Space Telescope Science Institute, 3700 San Martin Drive, Baltimore, MD 21218, USA}

\author[0000-0002-9820-1219]{Ekta~Patel}\thanks{Hubble Fellow}
\affiliation{Department of Astronomy, University of California, Berkeley, 501 Campbell Hall, Berkeley, CA, 94720, USA}
\affiliation{Department of Physics and Astronomy, University of Utah, 115 South 1400 East, Salt Lake City, Utah 84112, USA}

\author[0000-0001-8368-0221]{Sangmo Tony Sohn}
\affiliation{Space Telescope Science Institute, 3700 San Martin Drive, Baltimore, MD 21218, USA}

\author[0000-0003-4922-5131]{Andr\'es del Pino}
\affiliation{Centro de Estudios de F\'isica del Cosmos de Arag\'on (CEFCA), Unidad Asociada al CSIC, Plaza San Juan 1, 44001, Teruel, Spain}

\author[0000-0001-7827-7825]{Roeland P. van der Marel}
\affiliation{Space Telescope Science Institute, 3700 San Martin Drive, Baltimore, MD 21218, USA}
\affiliation{Center for Astrophysical Sciences, The William H. Miller III Department of Physics \& Astronomy, Johns Hopkins University, Baltimore, MD 21218, USA}

\author[0000-0001-9673-7397]{Mattia Libralato}
\affiliation{AURA for the European Space Agency (ESA), ESA Office, Space Telescope Science Institute, 3700 San Martin Drive, Baltimore, MD 21218, USA}

\author[0000-0002-1343-134X]{Laura L. Watkins}
\affiliation{AURA for the European Space Agency (ESA), ESA Office, Space Telescope Science Institute, 3700 San Martin Drive, Baltimore, MD 21218, USA}

\author[0000-0002-6054-0004]{Antonio Aparicio}
\affiliation{Intituto de Astrofísica de Canarias, 3200, La Laguna, Tenerife, Canary Islands, Spain}
\affiliation{Departamento de Astrofísica, Universidad de La Laguna, 3200, La Laguna, Tenerife, Canary Islands, Spain}

\author[0000-0003-0715-2173]{Gurtina Besla}
\affiliation{Steward Observatory, University of Arizona, 933 North Cherry Avenue, Tucson, AZ 85721, U.S.A.}

\author[0000-0001-6728-806X]{Carme Gallart}
\affiliation{Intituto de Astrofísica de Canarias, 3200, La Laguna, Tenerife, Canary Islands, Spain}
\affiliation{Departamento de Astrofísica, Universidad de La Laguna, 3200, La Laguna, Tenerife, Canary Islands, Spain}

\author[0000-0003-4207-3788]
{Mark A.\ Fardal}
\affiliation{Eureka Scientific, 2452 Delmer Street, Suite 100, Oakland, CA 94602, U.S.A.}

\author[0000-0001-5292-6380]{Matteo Monelli}
\affiliation{Intituto de Astrofísica de Canarias, 3200, La Laguna, Tenerife, Canary Islands, Spain}
\affiliation{Departamento de Astrofísica, Universidad de La Laguna, 3200, La Laguna, Tenerife, Canary Islands, Spain}
\affiliation{INAF-Osservatorio Astronomico di Roma, via Frascati 33, 00040 Monte Porzio Catone, Italy}

\author[0000-0001-5618-0109]{Elena Sacchi}
\affiliation{Leibniz-Institut für Astrophysik Potsdam, An der Sternwarte 16, 14482 Potsdam, Germany}

\author[0000-0002-9599-310X]{Erik Tollerud}
\affiliation{Space Telescope Science Institute, 3700 San Martin Drive, Baltimore, MD 21218, USA}

\author[0000-0002-6442-6030]{Daniel R. Weisz}
\affiliation{Department of Astronomy, University of California, Berkeley, 501 Campbell Hall, Berkeley, CA, 94720, USA}


\begin{abstract}
We have determined the proper motions (PMs) of 12 dwarf galaxies in the Local Group (LG), ranging from the outer Milky Way (MW) halo to the edge of the LG. We used HST as the first and Gaia as the second epoch using the GaiaHub software. For Leo A and Sag DIG we also used multi-epoch HST measurements relative to background galaxies. Orbital histories derived using these PMs show that two-thirds of the galaxies in our sample are on first infall with $>$90\% certainty. The observed star formation histories (SFHs) of these first-infall dwarfs are generally consistent with infalling dwarfs in simulations. The remaining four galaxies have crossed the virial radius of either the MW or M31. When we compare their star formation (SF) and orbital histories we find tentative agreement between the inferred pattern of SF with the timing of dynamical events in the orbital histories. For Leo~I, SF activity rises as the dwarf crosses the MW's virial radius, culminating in a burst of SF shortly before pericenter ($\approx1.7$~Gyr ago). The SF then declines after pericenter, but with some smaller bursts before its recent quenching ($\approx0.3$~Gyr ago). This shows that even small dwarfs like Leo~I can hold on to gas reservoirs and avoid quenching for several Gyrs after falling into their host, which is longer than generally found in simulations. Leo~II, NGC~6822, and IC~10 are also qualitatively consistent with this SF pattern in relation to their orbit, but more tentatively due to larger uncertainties.

\end{abstract}

\keywords{Proper motions (1295), Dwarf Galaxies (416), Local Group (929)}

\section{Introduction} \label{sec:intro}




Star formation and quenching are key issues in the study of dwarf galaxies. Within the Local Group (LG) most dwarf galaxies within the virial radius of the massive galaxies---Andromeda (M31) or the Milky Way (MW)---are quenched except for the most massive dwarfs such as the Magellanic Clouds. On the other hand, field dwarfs---those outside the virial radius of a massive galaxy---are generally gas-rich and star forming \citep[][]{Spekkens_2014,Putman_2021}. This trend is also seen in simulations of LG-like galaxy groups and isolated MW-like galaxies \citep[e.g.][]{Fattahi_2016,Sawala_2016,Wetzel_2016,Simpson_2018,Garrison-Kimmel2019,Libeskind_2020,Engler_2021,Engler_2023,Font_2021,Applebaum_2021,Akins_2021,Samuel_2022}{}{}. 

However, there are counter examples to this simple model. There are dwarf galaxies outside the influence of a massive galaxy that are quenched and gas-poor, such as Tucana \citep{Lavery_1992} or Cetus \citep{Whiting_1999}. These are theorized to be backsplash galaxies, dwarfs that were previously inside the hosts' virial radius, but are on elliptical orbits and now more distant, despite the previous quenching event \citep[e.g][]{Gill_2005,Sales_2007,Fraternali_2009,Santos_santos_2023}. 
Searches for dwarf galaxies beyond the LG around MW analog galaxies \citep[e.g.][]{Chiboucas_2013,Geha_2017,smercina_2018,Crnojevic_2019,Bennet_2019,Bennet_2020,Mao_2021,Carlsten_2022}{}{} have also yielded more star-forming dwarfs than expected from simulations or the LG, even when limited to only dwarfs projected inside the virial radius, the so-called `quenching problem' \citep[][]{Karunakaran_2021,Karunakaran_2023,Font_2022,Sales_2022, Jones2023}{}{}. 

Determining the orbits of dwarf galaxies within the LG is critical to potentially reconcile these issues. When data are precise enough, orbital histories allow a concrete determination of whether a dwarf galaxy is backsplash or on first-infall and, therefore, its expected star formation history (SFH) and quenching status, and how this compares to observations. 
It can also determine the length of time that it takes for dwarf galaxies to quench observationally by tracking how long the dwarfs are inside the hosts' virial radius before quenching occurs. 
Such orbital histories require full 6D position-velocity information, for which we require proper motions (PMs).  

PMs from the \textit{Gaia} mission \citep{Gaia_2016}, even with its interim data releases, have allowed a revolution in our understanding of the MW and its satellite system. The greater depth afforded by \textit{Gaia} compared to previous all-sky astrometry missions, such as Hipparcos, yields PMs out to the MW halo and beyond, including for its dwarf galaxy satellites. \textit{Gaia} has measured the PMs of almost all of the MW's globular clusters \citep[e.g.][]{gaia_2018,Baumgardt_2019,Vasiliev_2021}, the globular clusters of the Large Magellanic Cloud \citep{Bennet_2022}, and the dwarf satellites of the MW \citep[e.g.][]{Fritz_2018,McConnachie_2020b,Battaglia_2022}. Other key results from \textit{Gaia} including examining the internal structure of the MW \citep[e.g.][]{Antoja_2018}, including the discovery of the Gaia-Enceladus merger remnant \citep[][]{Belokurov_2018, Helmi_2018}, and the repeated impact of the Sagittarius dwarf galaxy on the star formation of the MW \citep[][]{Ruiz-Lara_2021}. 

However, as successful as \textit{Gaia} has been, the most recent \textit{Gaia} Data Release 3 (DR3) has an observation interval of only 34 months \citep[][]{Gaia_2022}. This limits the PM precision that can be obtained using \textit{Gaia} alone \footnote{The time baseline will increase in later data releases, up to a maximum possible mission length of around 10 years, but this is still not long enough for the kind of PM precision we need for this work.}. Nonetheless, \textit{Gaia} is also limited by the magnitude of the stars it observes; \textit{Gaia}'s astrometric errors rise rapidly from $G \sim 12$ down to its catalog limit of $G \sim 21$. 
This limits the distance to which \textit{Gaia} alone will be able to determine meaningful PMs. For example, a red giant star at the distance of the Andromeda galaxy (almost 800 kpc away) has a maximum observed magnitude of $G \approx 23$, far below the catalog cut-off for \textit{Gaia}. This means that \textit{Gaia} cannot measure the PMs of distant Local Group (LG) dwarf spheroidal galaxies that have exclusively old stellar populations. Instead, for dwarf irregular galaxies that have some recent star formation, \textit{Gaia} can at best detect some upper main sequence stars and/or supergiants. The small number and limited PM accuracy for such stars restricts the ability of \textit{Gaia} by itself to tightly constrain the PMs and orbits of such distant galaxies.  


Therefore, our team recently developed, validated, and released a software tool, \gaiahub\footnote{https://github.com/AndresdPM/GaiaHub}, to enable such PM measurements \citep{del_Pino_2022}. \textit{HST} has greater photometric sensitivity than \textit{Gaia}, allowing positions to be measured more accurately than by \textit{Gaia} for sources fainter than $G \approx 17.5$. Using \textit{HST} data as a first epoch and \textit{Gaia} as the second epoch then yields an increase in time baseline compared to \textit{Gaia} DR3.
This allows PM measurements with time baselines of up to $\sim$20 years, greatly increasing the precision compared to the 34 months of \textit{Gaia} alone. This expands on and automates the approach previously used in \cite{Massari_2020}, among others.

A second method to obtain accurate PMs for distant dwarf galaxies is to obtain two or more epochs of \textit{HST} data for a given target galaxy, spaced by many years. This is observationally expensive, and possible only for targets that pass a stringent time allocation process. However, once such data are available, they yield exquisite PM measurements \citep[e.g.,][]{Sohn_2013,Sohn_2017, sohn20}. In general \textit{HST} has the disadvantage of a small field of view, especially compared to the all-sky nature of \textit{Gaia}. However, this is not a significant issue for distant dwarf galaxies, which can be mostly covered with a single \textit{HST} pointing.   

In this paper, we report on the use of \gaiahub{} to determine the bulk PMs of 12 distant (D$>$200 kpc) Local Group dwarf galaxies. We demonstrate the general improvement possible compared to the use of \textit{Gaia} alone. Moreover, we report new multi-epoch \textit{HST} PM measurements for two of the sample galaxies. This yields enhanced accuracy, as well as a consistency check on the results derived using \gaiahub.  

The paper is organized as follows. 
In \S \ref{sec:PM}, we define the sample, derive new PMs, and compare them to existing literature PMs. In \S \ref{sec:orbits}, we use the newly-derived PMs, combined with literature positions, distances and line-of-sight (LOS) velocities, to derive possible orbital histories for our sample of LG dwarfs. In \S \ref{sec:disc}, we discuss the implications of these results. We compare the orbital and SFHs of the dwarfs. We also determine which dwarfs in the sample are potential backsplash galaxies and which are on first infall. Finally, in \S \ref{sec:Con}, we summarize our work. 

\begin{deluxetable*}{c|c|cc|cc|c}
\tablenum{1}
\tablecaption{Local Group Dwarfs Properties \label{tab:propr}}
\tablewidth{0pt}
\tablehead{
\colhead{Dwarf}  & \colhead{Dwarf} & \colhead{First Epoch} & \colhead{Time} &  \colhead{Distance} & \colhead{Distance} & \colhead{Line-of-sight}  \\
\multicolumn1c{Name} & \multicolumn1c{Type} & \colhead{HST} & \colhead{Baseline} & \colhead{Modulus} &  \colhead{(kpc)} & \colhead{Velocity}  \\
\multicolumn1c{} & \multicolumn1c{} &\multicolumn1c{Program ID} & \colhead{(Years)}  & \colhead{(mag)} &  \colhead{} & \colhead{(km $s^{-1}$)} } 
\startdata
IC~10 & dIrr & 9683 & 14.62 & 24.50$\pm$0.12 & 794$\pm$44 & -348.0$\pm$1.0 \\
IC~1613 & dIrr & 10505 & 10.77 & 24.39$\pm$0.12 & 755$\pm$42 & -233.0$\pm$1.0  \\
Leo~I & dSph & 10520 & 11.33 & 22.02$\pm$0.13 & 254$\pm$15 & 282.5$\pm$0.1  \\
Leo~II & dSph & 12304 & 5.18 & 21.84$\pm$0.13 & 233$\pm$14 & 78.0$\pm$0.1 \\ 
Leo A & dIrr & 10590 & 11.40 & 24.51$\pm$0.12 & 798$\pm$44 & 22.3$\pm$2.9  \\
Leo T & dTrans & 12914 & 4.37 & 23.10$\pm$0.10 & 417$\pm$19 & 38.1$\pm$2.0  \\
NGC~6822 & dIrr & 12180 & 6.60 & 23.31$\pm$0.08 & 459$\pm$17 & -57.0$\pm$2.0  \\
Peg dIrr & dIrr & 13768 & 1.84 & 24.82$\pm$0.07 & 920$\pm$30 & -183.0$\pm$5.0 \\
Phoenix & dTrans & 12304 & 5.32 & 23.09$\pm$0.10 & 415$\pm$19 & -13.0$\pm$9.0  \\
Sag DIG & dIrr & 9820 & 13.78 & 25.14$\pm$0.18 & 1067$\pm$88 & -78.5$\pm$1.0  \\
UGC 4879 & dIrr & 11584 & 7.26 & 25.67$\pm$0.04 & 1361$\pm$25 & -70$\pm$15.0  \\
WLM & dIrr & 12902 & 3.95 & 24.85$\pm$0.08 & 933$\pm$34 & -130.0$\pm$1.0  \\\enddata\label{tab:LOSv}
\tablecomments{All distances and line-of-sight velocities are from \cite{McConnachie_2012}. 
}
\end{deluxetable*}

\section{Proper Motion Analysis} \label{sec:PM}

\subsection{Sample Selection} \label{subsec:sample}

One of the goals of our project is to determine improved systemic PMs for a significant sample of distant LG dwarf galaxies. This requires new techniques or data compared to, e.g., the \textit{Gaia} DR3-based\footnote{Gaia Data Release 3 (DR3) and early data release 3 (eDR3) have the same PM results. Therefore to avoid confusion we will use DR3 throughout this work when referring to PM results from this data set, even if it relates to literature results that used eDR3.} results of \cite{Battaglia_2022}. Therefore, we limit our sample selection on those galaxies for which improvements are possible with \gaiahub. 

Our initial sample of LG dwarf galaxies was drawn from \cite{McConnachie_2012}, and then constrained further to use \gaiahub. First, we need dwarf galaxies that have suitable \textit{HST} images to be the first epoch of observations. To be suitable these images must meet a number of requirements. The \textit{HST} images must be from either the WFC3/UVIS or WFC/ACS instruments, as only they have sufficient resolution to correctly center the stars for such distant objects with the accuracy needed to determine PMs. The \textit{HST} images must also be in a wide-band filter that has a close color match to the \textit{Gaia} filters to enable confident cross matching of stars between the \textit{HST} and \textit{Gaia} data. In practice, this means F606W and F814W are preferred. Finally, the images need to be taken before May 26th, 2017 to add time baseline to the \textit{Gaia} DR3 observations. We find that images must be from before 2012 to add significant benefits over using \textit{Gaia} alone to determine PMs \citep[see][for more details]{del_Pino_2022}{}{}.

The second constraint on the sample is that the dwarf galaxies need to be far enough that \textit{Gaia} and \textit{HST} can work together to maximize the gains over \textit{Gaia} DR3. If the dwarf galaxies are too close to the observer, the brightest member stars will saturate the \textit{HST} imaging, significantly reducing the goodness of the fits. Simultaneously, these bright stars will be better fit by \textit{Gaia} DR3. Therefore, the benefits of using \gaiahub{} are only apparent for more distant objects\footnote{This is also due to the common observing strategies with HST for LG dwarf galaxies. With shorter HST observations \gaiahub{} operates well with closer sources such as MW GCs. }. For this sample, we impose a distance constraint of $D_{MW}>$ 200 kpc. 

Finally, there must be sufficient stars in common between the \textit{HST} and \textit{Gaia} catalog to make a PM fit. This means dwarf galaxies need to have enough member stars above the \textit{Gaia} catalog limit of $ G \sim 21$. The consequence for our sample is that for more distant systems (those with D$\gtrsim$300 kpc) our dwarfs are irregular or transitional, as fully quenched galaxies will not have enough bright member stars. 

These constraints leave us with 12 dwarfs in the LG where \gaiahub{} can offer a possible improvement over \textit{Gaia} DR3 PM results. The properties of these galaxies and the first-epoch \textit{HST} observations are listed in Table~\ref{tab:propr}. 

For two of the galaxies (Leo~A and Sag DIG), we also have multi-epoch \textit{HST} data from our programs GO-12273 (PI: van der Marel) and GO-15911 (PI: del Pino), as well as older archival data. In Section~\ref{subsec:HSTnew}, we use these to derive new multi-epoch \textit{HST} vs.~\textit{HST} PMs. These serve as an extra consistency check on the \gaiahub{} results for these two galaxies, and also help to further improve the accuracy of their orbital calculations. 

\begin{figure*}
    \centering
    \includegraphics[width=\linewidth]{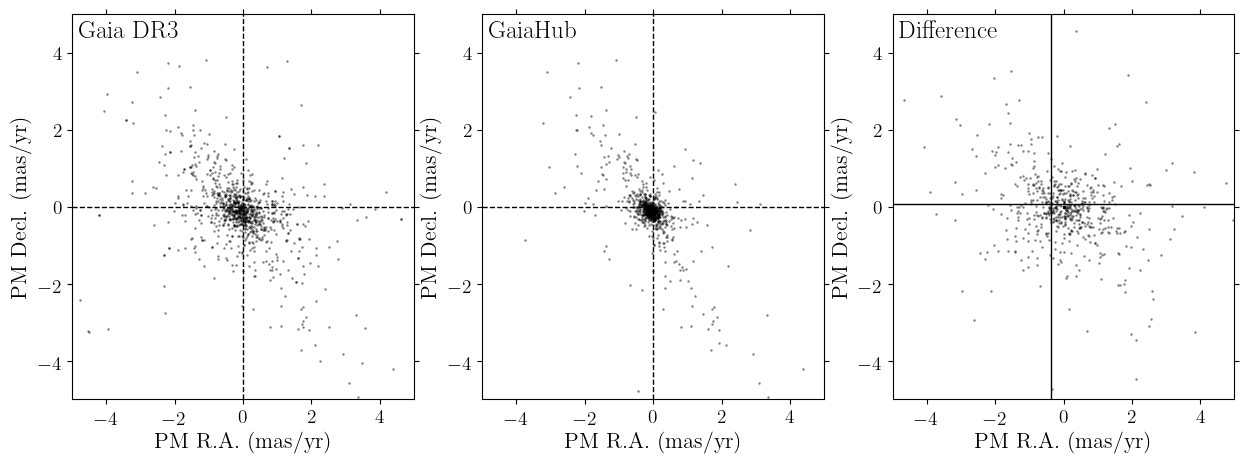}
    \caption{PM diagrams for individual sources in the \gaiahub{} field of view for Leo~I. Left: \textit{Gaia} DR3, Center: \gaiahub, Right: Difference between the measurements between \textit{Gaia} DR3 and \gaiahub. These are all sources that are found in the \textit{Gaia} DR3 and \gaiahub{} frame before any cleaning,  membership selection, or absolute astrometric alignment of the relative \gaiahub{} PMs. The black dashed lines are centered on zero PM in the left and center panels. In the right panel, the solid black lines show the weighted mean of the difference between the relative and absolute frames. Note the tighter clustering of sources in the \gaiahub{} frame. The diagonal feature seen in the left and center panels is likely the result of a systematic in Gaia where PM R.A. and PM Decl. is correlated in some sources.  }
    \label{figure:PM_LeoI}
\end{figure*}

\subsection{GaiaHub Analysis} \label{subsec:GH}

\subsubsection{Absolute Astrometric Alignment} \label{subsubsec:frame}


With our sample selected, we first ran \gaiahub{} on these dwarfs using the prescription from \cite{del_Pino_2022}. 
In normal use, \gaiahub{} would use only the member stars to align the HST epoch with \textit{Gaia}. This procedure is iterative, selecting member stars in the VPD and repeating the epoch alignment until convergence. Once the relative PMs are determined, \gaiahub{} will compute the absolute PMs by adding the error-weighted mean difference between the relative PMs and the absolute ones measured by Gaia. However, this process was optimized and tested for dense fields, such as those of globular clusters or nearby dwarf galaxies, with a large number of sources and a well-populated foreground population. This is not the case for our sample of galaxies, where member counts are low and the foreground population is even lower. Therefore, we chose to not use the automatic membership selection method implemented in \gaiahub{} nor the absolute reference frame. Instead, we use \gaiahub{} {\it manually}, forcing it to use all the stars in the field to perform the epoch alignment in just one iteration.

This more manual process starts by taking the \textit{Gaia} DR3 absolute PMs and comparing them to the relative \gaiahub{} PMs. The target dwarf galaxy is generally apparent as an overdensity in PM space in both frames. We use Leo~I to illustrate the process and show the corresponding PM diagrams in Figure \ref{figure:PM_LeoI} (left and center panels, respectively). The value of the \gaiahub{} PMs is immediately apparent, as they form a tighter clump in PM space than the DR3 PMs. However, they need to be placed on an absolute frame to be used in dynamical calculations. 

First, we manually shift (i.e., add an initial constant PM offset to) the \gaiahub{} PMs so that the centroid of its overdensity peak roughly matches that in the DR3 data. We then compare the PMs of individual sources (right panel of Figure \ref{figure:PM_LeoI}). If a source has a PM that does not match between DR3 and \gaiahub{} at the 3$\sigma$ level, then this source is removed from the alignment process as it likely has a poor measurement in either the \textit{Gaia} DR3 catalog, the \gaiahub{} relative frame, or both. This could be caused by a number of factors, the investigation of which is beyond the scope of this work. 

\begin{figure*}
    \centering
    \includegraphics[width=0.71\linewidth]{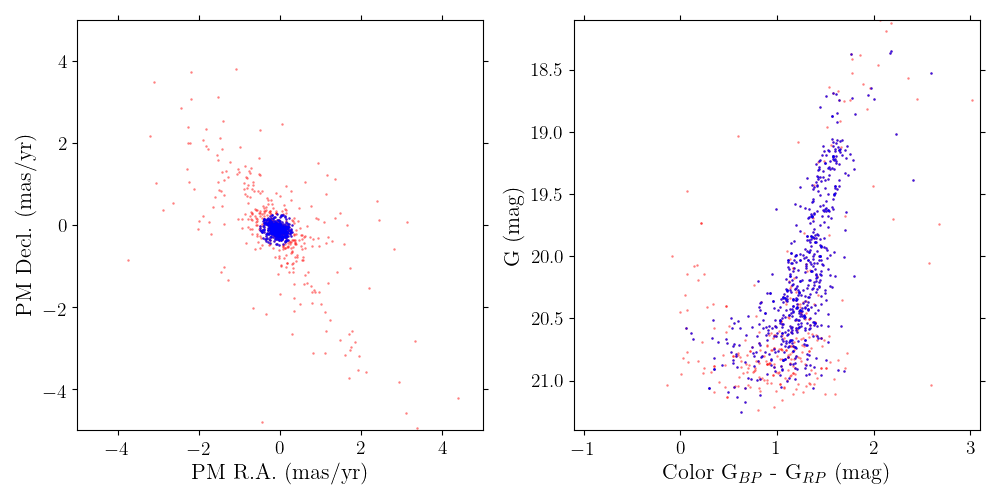}
    \caption{The final member selection for Leo~I after frame matching and the iterative membership selection. The member stars are shown in blue and the non-members in red. Left: The relative PMs from \gaiahub{} frame. Right: The CMD from the \textit{Gaia} DR3 catalog.}
    \label{figure:CMD_LeoI}
\end{figure*}

\begin{figure*}
    \centering
    \includegraphics[width=0.71\linewidth]{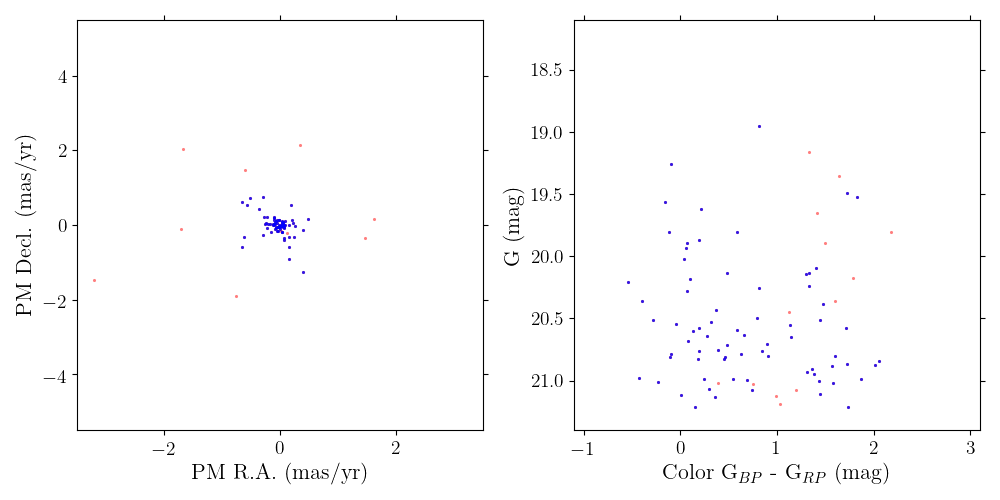}
    \includegraphics[width=0.71\linewidth]{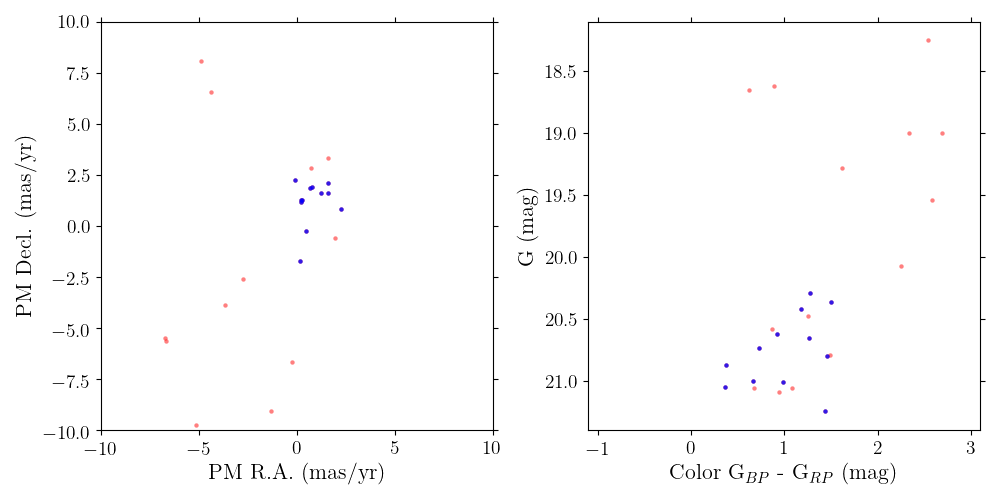}

    \caption{The same as Figure \ref{figure:CMD_LeoI} but for Leo A (Upper Panels), which represents a typical result, and Leo T (Lower Panels), which represents our least certain result.}
    \label{figure:CMD_other}
\end{figure*}

Once the bad matches have been removed, we shift the \gaiahub{} PMs so that they best match the DR3 PMs for the remaining sources. This is performed using an iterative weighted-mean approach. First the weighted mean of both frames is calculated, then the relative frame is shifted so that the weighted mean for both frames is the same. Once this is done, the test for poor matches is repeated and any that are found are removed from the sample. This process is then repeated until no sources are removed. In practice this process is very quick; few dwarfs in our sample need more than one iteration and none need more than three. The random uncertainty in the absolute PM calibration of the \gaiahub{} sources is also calculated and stored for later reference.

Once the \gaiahub{} astrometric alignment has been completed, we once again consider all measured sources, including those without corresponding DR3 PMs. These are stars that have positions in both the \textit{Gaia} and \textit{HST} data but not PMs in the \textit{Gaia} data, generally because they are too faint. 

\subsubsection{Membership selection} \label{subsubsec:member}

Once the \gaiahub{} PMs are on an absolute frame, it is important to select the member stars of the target galaxy to determine its bulk PM. The membership selection is independent of the absolute bulk PM,
and it is therefore performed in the relative frame.

We fit the relative PM distribution with a 2D Gaussian profile centered on the weighted mean. All stars with PMs more than 3$\sigma$ from the centroid are removed. Then the weighted average and weighted standard deviation are recalculated. This iterates until no further stars are removed. The left panel of Figure~\ref{figure:CMD_LeoI} shows the result for Leo~I. Once this PM selection has been completed, the Color-Magnitude Diagram (CMD) is examined to determine if the member stars resemble the expected distribution for the galaxy. If any stars are significant outliers from the expected color and magnitude for the dwarf, they are removed as member stars. In practice, most foreground stars are removed based on their PMs, and this final CMD step only removes prospective members for a few dwarfs and the changes are $<$1\% of possible members. 

The right panel of Figure~\ref{figure:CMD_LeoI} shows the CMD for Leo~I. It reveals predominantly red giant branch (RGB) stars with a tip around \textit{Gaia} G$\sim$19~mag. We note that Leo~I is one of the galaxies for which we have the largest number of stars. The top panel of Figure~\ref{figure:CMD_other} is similar to 
Figure~\ref{figure:CMD_LeoI}, but for the case of Leo~A, which represents a more typical galaxy for our sample. Note that the CMD reveals a significant young/intermediate-age population at the blue end. The bottom panel of Figure~\ref{figure:CMD_other} shows the results for Leo~T, which represents the sparsest galaxy in the sample. 


\subsubsection{Proper motion results} \label{subsubsec:GH_results}

We explored and compared two methods for calculating the systemic PM of each target galaxy. In both methods, we only use member stars determined by \gaiahub{} as described in Section~\ref{subsubsec:member}. In Method~1, we calculate the weighted average of the \gaiahub{} PMs. The uncertainty in the systemic PM is the quadrature sum of the uncertainty in the weighted average, and the uncertainty in the astrometric PM calibration (from Section~\ref{subsubsec:frame}). In Method~2, we calculate the weighted average of the \textit{Gaia} DR3 PMs of the member stars, and its corresponding uncertainty. 

The \gaiahub{} PMs are generally more accurate than the \textit{Gaia} DR3 PMs, due to our selection of target galaxies with large \textit{HST} vs.~\textit{Gaia} time baselines (see Section~\ref{subsec:sample}). Thus, Method one is preferred as long as the number of stars in the field is sufficiently large. In fields with low star density, where \gaiahub{} can introduce systematic errors, Method 2 is preferred. In line with these considerations, we adopted Method~1 for IC~1613, Leo~I \& Leo~II, and Method~2 for the remaining galaxies. 

Where multiple \textit{HST} epochs are available for the same pointing, we use the oldest epoch in order to maximize the time baseline. Where multiple \textit{HST} pointings exist for a single dwarf we use all available pointings, again with the oldest epoch possible, to maximize the number of member stars. Each pointing is treated independently and the final systemic PMs are averaged between all available pointings. 
The final inferred systemic PMs are listed with their uncertainties in Table~\ref{tab:PM}. 

\begin{deluxetable*}{c|cc|c|cc|c}[h]
\tablenum{2}
\tablecaption{Local Group Dwarf Galaxy Proper Motions \label{tab:PM}}
\tablewidth{0pt}
\tablehead{
\colhead{Dwarf}  & \colhead{This work} & \colhead{This work} &  \colhead{Method} & \colhead{Literature} & \colhead{Literature} & \colhead{Source} \\
\multicolumn1c{Galaxy} & \multicolumn1c{$\mu^{*}_{\alpha}$} & \multicolumn1c{$\mu_{\delta}$} & \multicolumn1c{Used} &  \multicolumn1c{$\mu^{*}_{\alpha}$} & \multicolumn1c{$\mu_{\delta}$} & \multicolumn1c{ }  \\
\multicolumn1c{Name} & \multicolumn1c{(mas yr$^{-1}$)} & \multicolumn1c{(mas yr$^{-1}$)} & \multicolumn1c{ }  & \multicolumn1c{(mas yr$^{-1}$)} & \multicolumn1c{(mas yr$^{-1}$)} & \multicolumn1c{ }  }
\startdata
IC~10 & 0.002$\pm$0.027 & 0.012$\pm$0.037 & Method 2 & -0.002 $\pm$0.008 & 0.020$\pm$0.008 & \citet{Brunthaler_2007} \\
IC~1613 &  0.030$\pm$0.012 & 0.001$\pm$0.011 & Method 1 &  0.04$\pm$0.02 & 0.01$\pm$0.01 & \cite{Battaglia_2022} \\
Leo~I &  -0.063$\pm$0.005 & -0.111$\pm$0.004 & Method 1 &   -0.06$\pm$0.01 & -0.12$\pm$0.01 & \cite{Battaglia_2022} \\
Leo~II & -0.143$\pm$0.009 & -0.130$\pm$0.009 & Method 1 &  -0.11$\pm$0.03 & -0.14$\pm$0.03 & \cite{Battaglia_2022} \\ 
Leo A & -0.066$\pm$0.073 & -0.044$\pm$0.087 & Method 2 &  -0.06$\pm$0.09 & -0.06$\pm$0.09 & \cite{Battaglia_2022} \\
Leo T & -0.007$\pm$0.242 & 0.110$\pm$0.168 & Method 2 & 0.23$\pm$0.37 & -0.12$\pm$0.22 & \cite{Battaglia_2022} \\
NGC~6822 &   -0.003$\pm$0.049 & -0.087$\pm$0.050 & Method 2 &  -0.06$\pm$0.01 & -0.07$\pm$0.01 & \cite{Battaglia_2022} \\
Peg dIrr & 0.173$\pm$0.217 & 0.041$\pm$0.142 & Method 2 & 0.15$\pm$0.14 & 0.07$\pm$0.12 & \cite{Battaglia_2022} \\
Phoenix & 0.142$\pm$0.028 & -0.040$\pm$0.032 & Method 2 &  0.08$\pm$0.03 & -0.06$\pm$0.04 & \cite{Battaglia_2022} \\
Sag DIG & 0.059$\pm$0.101 & -0.128$\pm$0.092 & Method 2 & 0.11$\pm$0.19 & -0.37$\pm$0.17 & \cite{Battaglia_2022} \\
UGC 4879 & 0.019$\pm$0.110 & -0.011$\pm$0.067 & Method 2 & 0.00$\pm$0.11 & -0.04$\pm$0.09 & \cite{Battaglia_2022} \\
WLM &  0.116$\pm$0.081 & -0.056$\pm$0.069 & Method 2 & 0.09$\pm$0.03 & -0.07$\pm$0.02 & \cite{Battaglia_2022} \\
\enddata
\tablecomments{$\mu^{*}_{\alpha}$ = $\mu_{\alpha} * cos(\delta) $. \cite{Battaglia_2022} PMs are based on \textit{Gaia} DR3 data. \cite{Brunthaler_2007} PM is based on the motion of an H2O maser. For the latter result we included their correction for the rotation of IC~10, but to get an absolute PM, we did not include their correction for the solar reflex motion. A visual comparison between these PM measurements can be seen in Figure \ref{figure:Error_comp}}
\end{deluxetable*}

\begin{figure*}
    \centering
    \includegraphics[width=\linewidth]{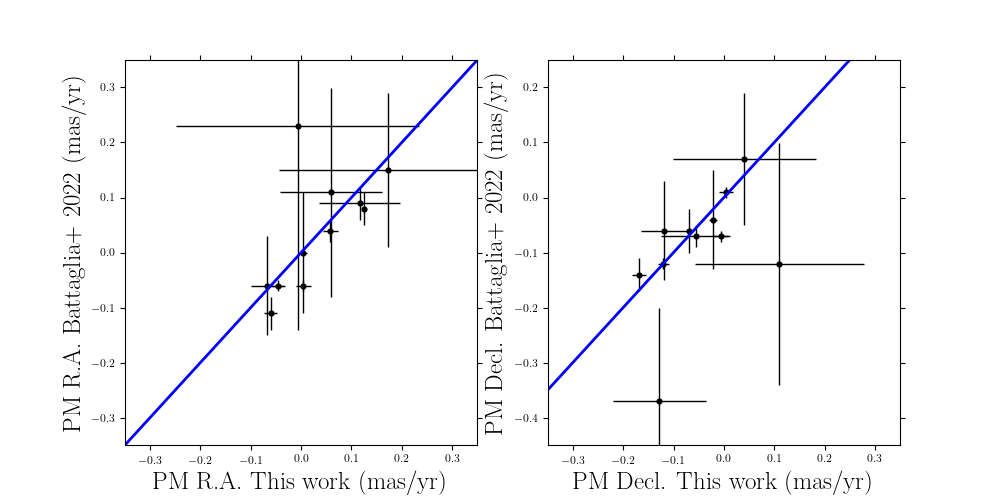}

    \caption{A comparison of the systemic PMs for our dwarf galaxy sample between this work and \citet{Battaglia_2022}, for each coordinate on the sky.  The results are in adequate agreement. Our new uncertainties are generally smaller. This plot does not include our HST-HST PM results or the IC~10 PM from \cite{Brunthaler_2007}.}
    \label{figure:Error_comp}
\end{figure*}


\subsubsection{Comparison to \textit{Gaia} DR3 Results\label{subsubsec:litGaia_pm}}

To validate our PM results it is useful to compare them to independent measurements. 
First we compare to the PMs inferred from \textit{Gaia} DR3 by \cite{Battaglia_2022}, as listed in Table~\ref{tab:PM}. The comparisons are shown in Figure~\ref{figure:Error_comp}. The top panel of Figure~\ref{figure:Error_comp} shows that the measurements are generally in agreement to within 1$\sigma$. The largest differences are when the measurement uncertainties are large, as might be expected. The $\chi^2$ of the comparison is $9.3$ for $N=22$ degrees of freedom. For uncorrelated measurements one would expect $\chi^2 \approx N \pm \sqrt{2N} = 22.0 \pm 6.6$. The fact that we find a lower value is not unexpected, since our results rely in part on the same underlying data as those from \cite{Battaglia_2022}, so that the measurements are not uncorrelated.

The uncertainties in our new PMs are generally smaller than those in \cite{Battaglia_2022}. This is most pronounced for the galaxies for which we used Method 1 in Section~\ref{subsubsec:GH_results}: the median and average uncertainty for our results are, respectively,
0.009 and 0.0083 mas/yr
versus 
0.015 and 0.018 mas/yr
for \cite{Battaglia_2022}. There is a minor improvement for Method~2 galaxies as well: 
0.084 and 0.100 mas/yr
for our work versus 
0.090 and 0.108 mas/yr
for \cite{Battaglia_2022}. 

For Method~2, both measurements are based on averages of \textit{Gaia} DR3 PMs. However, the sizes of stellar samples that are averaged differ between our studies.
For three dwarfs (NGC~6822, Peg dIrr and WLM) we obtain larger uncertainties than those obtained by \cite{Battaglia_2022}. This is likely because the angular size of the dwarfs on the sky is such that large parts of the dwarfs are outside the \textit{HST} FOV, as well as the relatively short time baseline in the case of Peg dIrr and WLM. As a result, more stars can be found with the all-sky coverage of \textit{Gaia}. However, for other Method~2 galaxies, our uncertainties are smaller, despite the smaller field. This improvement is mainly possible because we can make a cleaner membership selection using the relative \gaiahub{} measurements. This allows more stars to be securely identified as members, which reduces the PM uncertainties (which scale as $1/\sqrt(N)$, where $N$ the number of stars). Moreover, it reduces the number of remaining foreground contaminant stars. The latter yields lower systematic uncertainties, even for those galaxies where the random uncertainties of \cite{Battaglia_2022} may be lower than ours. 

\subsubsection{Comparison to Other Literature Results\label{subsubsec:lit_pm}}

For some of our target galaxies there exist PM measurements other than from \textit{Gaia} DR3. These provide additional validations of our PM results. If a sample galaxy is not listed, it has no literature PM measurements aside from \textit{Gaia}. 

\bigskip

{\bf IC~10:} A PM measurement has been published based on radio observations of a water maser \citep{Brunthaler_2007}. This value is listed in the top right of Table~\ref{tab:PM}. This measurement allows a truly independent check of our methodology, as this PM measurement does not rely on \textit{Gaia} or \textit{HST} and is thus free of any of the systematics that could affect these telescopes. The uncertainties in the radio measurement are about four times smaller than ours. The measurements are in agreement to within these uncertainties. This excellent agreement limits the possible presence of any remaining systematics in our results.  

\bigskip

\begin{figure}
    \centering
    \includegraphics[width=\linewidth]{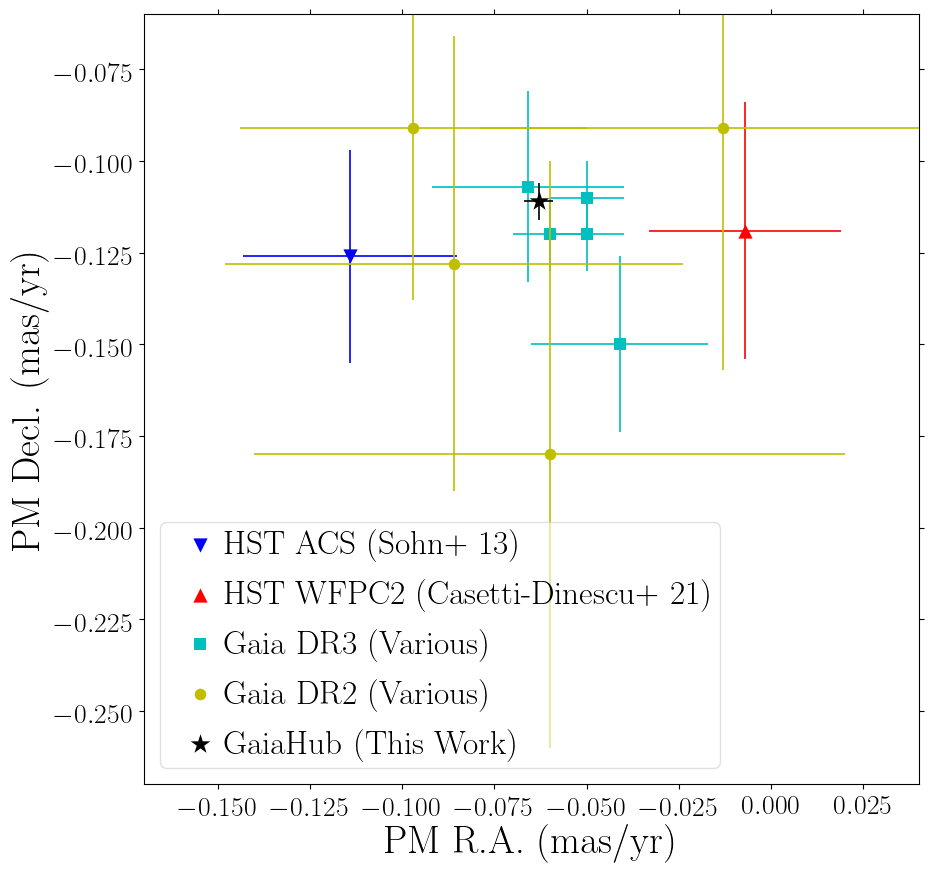}
    \caption{Comparison of the absolute PM measurements for Leo~I. The PM reported in this work is consistent though more precise than previous estimates \textit{HST}  ACS from \cite{Sohn_2013}. \textit{HST}  WFPC2 from \cite{Casetti_2022}. \textit{Gaia}  DR2 from \cite{gaia_2018, Fritz_2018, Simon_2018, McConnachie_2020a}. \textit{Gaia}  DR3 from \cite{McConnachie_2020b, Li_2021, Martinez_2021, Battaglia_2022,Pace_2022}.
    The different values reported using \textit{Gaia}  data are primarily the result of the different membership selection methods used by those works.}
    \label{figure:LeoI}
\end{figure}

{\bf Leo~I:} A number of different studies have determined the PM of Leo~I, and these results are shown in Figure \ref{figure:LeoI}. The uncertainties in our \gaiahub{} measurement are smaller than those in any of the existing literature measurements. The overall agreement between our result and the various existing \textit{Gaia} measurements is good, consistent with the analysis in Section~\ref{subsubsec:litGaia_pm}.
\cite{Sohn_2013} presented an \textit{HST}-only measurement using ACS images as both the first and second epoch. \cite{Casetti_2022} more recently presented a measurement using WFPC2 images as both the first and second epoch. The former paper used background galaxies to set the absolute reference frame while the latter paper uses astrometry of \textit{Gaia} sources. The uncertainties in both measurements are about six times larger than ours. The results are marginally consistent with our result, but both are offset (in opposite directions) in $\mu_\alpha^{*}$ by $\sim 2 \sigma$. 

\bigskip

{\bf Leo~II:} There is an existing two-epoch \textit{HST} WFC3 PM measurement from \cite{Piatek_2016}, using a background QSO to set the absolute reference frame. The uncertainty in the measurement is about four times larger than ours. It is marginally consistent with our result but also offset in $\mu_\alpha^{*}$ by $\sim 2 \sigma$. 

\subsection{New Multi-Epoch Hubble Measurements} \label{subsec:HSTnew}


%
\begin{deluxetable*}{lcccccccc}
\tablenum{3}
\tablecaption{Observation summary of Leo~A and Sag~DIG
              \label{t:obslog}}
\tablehead{
\colhead{}       & \multicolumn{2}{c}{{\bf Epoch~1}}    & \colhead{} & \multicolumn{2}{c}{{\bf Epoch~2}}    & \colhead{} & \multicolumn{2}{c}{{\bf Epoch~3}}   \\
\cline{2-3}\cline{5-6}\cline{8-9}
\colhead{Galaxy}       & \colhead{Date}    & \colhead{Exp. Time}      & \colhead{} & \colhead{Date}    & \colhead{Exp. Time}     & \colhead{} & \colhead{Date}    & \colhead{Exp. Time}
          }
\startdata
{\bf Leo~A}     & 2005-12-26 & 1220s$\times$16 & & 2011-12-28 & 1330s$\times$\phn8  & & 2020-12-17 & 1277s$\times$\phn8 \\
{\bf Sag~DIG}   & 2011-09-20 & 1317s$\times$12 & & 2020-09-06 & 1262s$\times$16     & & \nodata    & \nodata \\
\enddata
\tablenotetext{a}{Exposure times in (seconds) $\times$ (number of exposures) format. 
                  Here we list the average of the entire exposures, 
                  but the actual individual exposure times vary by 
                  only a few percent in length.}
\end{deluxetable*}
%


We have also used multi-epoch \hst\ data to obtain more accurate PM estimates for Leo~A and Sag~DIG. All images used for the PM measurements of these two dwarf galaxies were obtained with ACS/WFC using the F814W filter. Details of the epochs of observations and individual exposures are in Table~\ref{t:obslog}. In summary, we used three epochs of data with a total time baseline of 15 years for Leo~A, whereas only two epochs were used with a time baseline of 9 years for Sag~DIG. The 2011 data are from the \textit{HST} program GO-12273 (PI: van der Marel) and the 2020 data are from program GO-15911 (PI: del Pino).

To measure the PMs of Leo~A and Sag~DIG, we closely follow the established methods used by our previous work on M31, Leo~I, the Draco/Sculptor dwarf spheroidal galaxies, and NGC~147 and NGC~185 \citep{Sohn_2012, Sohn_2013, Sohn_2017, Sohn_2020}. All data were downloaded from the Mikulski Archive for Space Telescopes (MAST). We specifically downloaded the {\tt \_flc.fits} images that have been processed for the imperfect charge transfer efficiency (CTE) using the correction algorithms of \citet{Anderson_2010}. For each star, we determine a position and a flux from the {\tt \_flc.fits} images using the FORTRAN code {\tt hst1pass} \citep{Anderson_2022}. The measured positions are then corrected for geometric distortion using the solutions provided by \citet{Anderson_2006}. For each dwarf galaxy, we create high-resolution stacked images with pixel scales of $0.025$ mas using images from the reference epoch, i.e., the first-epoch (2005) data for Leo~A and the second-epoch (2020) data for Sag~DIG.

We identify stars associated with Leo~A and Sag~DIG via CMDs constructed from multi-band photometry. To construct CMDs, we used the F475W images obtained during the first-epoch observations for Leo~A, and F475W images obtained in 2003 observations for Sag~DIG.\footnote{While F814W images were also obtained in 2003 for the Sag~DIG field, they were too shallow to be considered as a separate epoch of observation for the PM measurements.} Background galaxies were identified first through an objective selection based on the quality-of-fit parameter output from the {\tt hst1pass} code, and then by visually inspecting each source. We thus identified 180 and 123 background galaxies that are suitable for defining a stationary astrometric reference frame, in the respective Leo~A and Sag~DIG fields. 

In the next step, we constructed an empirical ``template'' for each star and background galaxy by supersampling the scene extracted from the high-resolution stack. Each template constructed this way takes into account the point-spread function (PSF), the galaxy morphology (in case of the background galaxies), and the pixel binning. These templates are fitted on each individual {\tt \_flc.fits} image for measuring the position of each star or galaxy. For the reference epoch, we fitted the templates directly onto the individual images since the templates were constructed {\it from} that epoch. For the other non-reference epoch(s), we employed $7\times7$ pixel convolution kernels when fitting templates to allow for differences in PSF between different epochs. These convolution kernels were derived by comparing PSFs of numerous bright and isolated stars associated with the target dwarf galaxies between the reference and other epochs. At this stage, we were left with template-fitted positions of all stars and background galaxies in all individual images in each epoch.

We defined the reference frame for each LG dwarf galaxy (the frame with respect to which any motions are measured) by averaging the template-based positions of stars from repeated exposures of the reference epoch. The positions of stars in each exposure of the other epochs were then used to transform the template-measured positions of the galaxies into the reference frames. Subsequently, we measured the positional difference for each background galaxy between the reference and other epochs, relative to the stars associated with Leo~A and Sag~DIG. Additional corrections (a.k.a. ``local corrections'') were applied to correct for residual CTE and any remaining geometric distortion systematics, by making measurements only relative to stars with similar brightness in the local vicinity on the image.

Once we had the reflex displacements of background galaxies measured for each non-reference epoch, we calculated the final PMs as follows. For Sag~DIG, with only two epochs of observations, the PM calculations were carried out in a straightforward way following those in \citet{Sohn_2020}. Briefly, we took the error-weighted average of all displacements of background galaxies with respect to Sag~DIG stars for each individual first-epoch exposure to obtain an independent PM estimate. The PM and associated uncertainty of Sag~DIG were then obtained by the error-weighted mean of these individual PM estimates.

\begin{deluxetable}{c|cc}[t]
\tablenum{4}
\tablecaption{New Multi-Epoch HST Proper Motions \label{tab:PMHub}}
\tablewidth{0pt}
\tablehead{
\colhead{Dwarf}  & \colhead{HST} & \colhead{HST} \\
\multicolumn1c{Galaxy} & \multicolumn1c{$\mu^{*}_{\alpha}$} & \multicolumn1c{$\mu_{\delta}$} \\ 
\multicolumn1c{Name} & \multicolumn1c{(mas $yr^{-1}$)} & \multicolumn1c{(mas $yr^{-1}$)} }
\startdata
Leo A & 0.0019$\pm$0.0097 & -0.0836$\pm$0.0090 \\
Sag DIG & 0.0433$\pm$0.0166 & 0.0465$\pm$0.0191 \\
\enddata
\tablecomments{$\mu^{*}_{\alpha}$ = $\mu_{\alpha} * cos(\delta) $.}
\end{deluxetable}

For Leo~A, we instead first calculated the average position of each background galaxy in each epoch. This was done by directly adopting the averaged positions for the first (reference) epoch, and by converting the local-corrected displacements into positions for the second and third epochs. Because we have multiple measurements in each epoch, the mean position of each background galaxy in each epoch has a known associated uncertainty. We fitted straight lines through the X and Y positions as a function of $\Delta T$, being the time since the first epoch in units of years, using chi-square minimization of the residuals weighted with the measurement uncertainties. The resulting slope of this fitting process is the PM of each background galaxy. The PM and associated uncertainty of Leo~A were then obtained by the error-weighted mean of these individual PM estimates.

\begin{deluxetable*}{lcccccc}[h]
\tablenum{5}
\tablecaption{Local Group Dwarfs Position and Velocity Vectors \label{tab:Pos_vel}}
\tablewidth{0pt}
\tablehead{
\colhead{Dwarf}  &  \colhead{$x$} & \colhead{$y$} & \colhead{$z$} &  \colhead{$v_x$} & \colhead{$v_y$} & \colhead{$v_z$} \\
\multicolumn1c{} & \colhead{(kpc)} &  \colhead{(kpc)} & \colhead{(kpc)} &  \multicolumn1c{(km $s^{-1}$)} & \multicolumn1c{(km $s^{-1}$)}  & \multicolumn1c{(km $s^{-1}$)} 
}
\startdata
IC~10 & -392.0 $\pm$ 21.2 & 693.6 $\pm$ 38.4 & -45.1 $\pm$ 2.8 & 177.8 $\pm$ 26.4 & -55.3 $\pm$ 15.7 & 103.1 $\pm$ 30.5 \\
IC~1613 & -246.9 $\pm$ 13.2 & 285.2 $\pm$ 15.9 & -657.0 $\pm$ 36.6 & -3.5 $\pm$ 40.4 & 99.5 $\pm$ 38.4 & 218.1 $\pm$ 19.5 \\
Leo~I & -123.2 $\pm$ 6.8 & -119.6 $\pm$ 7.1 & 192.3 $\pm$ 11.3 & -122.8 $\pm$ 5.3 & -15.6 $\pm$ 10.2 & 137.3 $\pm$ 6.2 \\
Leo~II & -76.5 $\pm$ 4.1 & -58.2 $\pm$ 3.5 & 215.0 $\pm$ 12.9 & -89.3 $\pm$ 10.7 & 42.6 $\pm$ 15.5 & 4.9 $\pm$ 5.9\\
Leo A & -472.1 $\pm$ 34.9 & -141.5 $\pm$ 8.1 & 633.6 $\pm$ 25.5 & 72.9 $\pm$ 169.2 & -65.8 $\pm$ 324.6 & 10.3 $\pm$ 227.7 \\
Leo T & -254.9 $\pm$ 11.3 & -172.4 $\pm$ 7.9 & 288.5 $\pm$ 13.1 & -90.2 $\pm$ 368.9 & 426.4 $\pm$ 304.6 & 82.7 $\pm$ 334.0 \\
NGC~6822 & 385.1 $\pm$ 14.6 & 186.4 $\pm$ 6.9 & -145.9 $\pm$ 5.4 & 18.6 $\pm$ 55.2 & 54.3 $\pm$ 99.8 & -42.1 $\pm$ 101.7 \\
Peg dIrr & -65.3 $\pm$ 1.8 & 664.4 $\pm$ 21.7 & -633.8 $\pm$ 20.7 & -721.1 $\pm$ 879.5 & -64.0 $\pm$ 506.0 & 15.0 $\pm$ 503.4 \\
Phoenix & -3.5 $\pm$ 0.3 & -149.0 $\pm$ 6.8 & -387.3 $\pm$ 17.7 & -143.5 $\pm$ 59.3 & 21.4 $\pm$ 57.1 & 106.1 $\pm$ 23.5\\
Sag DIG & 946.9 $\pm$ 78.8 & 367.9 $\pm$ 30.3 & -301.7 $\pm$ 24.7 & -190.5 $\pm$ 222.9 & 488.7 $\pm$ 444.9 & -61.9 $\pm$ 487.6 \\
UGC 4879 & -967.5 $\pm$ 17.7 & 263.7 $\pm$ 4.8 & 928.7 $\pm$ 17.0 & 138.4 $\pm$ 496.6 & 163.8 $\pm$ 425.2 & 58.1 $\pm$ 513.7 \\
WLM & 53.8 $\pm$ 2.3 & 255.1 $\pm$ 9.3 & -895.3 $\pm$ 32.6 & -324.4 $\pm$ 346.2 & -231.0 $\pm$ 306.8 & -15.9 $\pm$ 93.6 \\
\enddata
\tablecomments{Three-dimensional position and velocity vectors derived, in Galactocentric coordinates, from the galaxy positions, distances, LOS velocities (see Table \ref{tab:LOSv}) and PM measurements. For IC~10 we used the water maser PM listed in Table \ref{tab:PM}. For Leo A and the Sag DIG we used the \textit{HST} PMs from Table \ref{tab:PMHub} For the remaining galaxies we used our newly derived PMs listed in Table \ref{tab:PM}. The uncertainties in this table are highly asymmetric as 3 (distance, PMs) of the observed 6 coordinates have much larger errors than the others (position, LOS velocity). }
\end{deluxetable*}

Finally, PMs of Sag~DIG and Leo~A were transformed to $\masyr$ via multiplying by the pixel scale of our reference frames (0.05~arcsec\,pix$^{-1}$) and by dividing by the time baseline of our observations. Also, we rotated the results in the reference frame to match the sky coordinates, and multiplied by $-1$ to infer the actual PM of the foreground LG dwarf galaxies from the reflex motion of the background galaxies. Final PM results are listed in Table~\ref{tab:PMHub}.

The uncertainties in our new measurements are 5--8 times smaller than those in the measurements obtained with the help of \gaiahub{} and listed in Table~\ref{tab:PM}. Only the $\mu_\delta$ result for the Sag DIG differs by more than $1\sigma$ ($1.9\sigma$ to be precise) between the measurements. However, this can be plausibly attributed to random variations. The $\chi^2$ of the comparison for all $N=4$ coordinates (2 coordinates for 2 galaxies) is $\chi^2 = 4.4$, where $\chi^2 \approx N \pm \sqrt{2N} = 4.0 \pm 2.8$ is expected as a result of random errors alone.



\section{Orbits of the LG Dwarfs} 
\label{sec:orbits}

\subsection{Orbit Methodology}

The galaxy distances, LOS velocities, and PM measurements were used to calculate the 6D position-velocity information in Table \ref{tab:Pos_vel}. For IC~10 we used the water maser PM from \citet{Brunthaler_2007} listed in Table \ref{tab:PM}. For Leo A and the Sag DIG we used the \textit{HST} PMs from Table \ref{tab:PMHub}. For the remaining galaxies, we use our newly derived \gaiahub{} PMs listed in Table \ref{tab:PM}.

To reconstruct orbits for the LG dwarf galaxies in our sample, we integrate the equations of motion backward in time using the 6D position-velocity information listed in Table \ref{tab:Pos_vel} as inputs into the orbital models. We follow the orbital methodology described in Section 3 of \citet{patel17a}. These methods are briefly described below.

For each galaxy in our sample, we compute a five-body orbit accounting for the combined gravitational potentials of the galaxy itself and the MW, M31, the Large Magellanic Cloud (LMC), and M33. The MW and M31 potentials are each modeled with three components: an NFW halo \citep{nfw96}, a Miyamoto-Nagai disk \citep{mn75}, and a Hernquist bulge \citep{hernquist90}. We adopt virial masses of $1\times 10^{12}\, M_{\odot}$ for the MW and $2\times 10^{12}\, M_{\odot}$ for M31 as in \citet{patel17a}. These masses are consistent with compilations of mass estimates for both galaxies provided in \citet{wang20, patel23}. We note that previous works, such as \citet{patel20}, have shown that even a factor of two difference in the mass of the MW can result in substantially different estimated orbital histories for lower mass satellites in their vicinity. As we show later in this section, a majority of our galaxies are on first infall and are still well outside the virial radii of the MW or M31 at present day, thus these galaxies are the least likely to be affected by a factor of two difference in the masses of the MW or M31 mass. The galaxies most likely to be affected as they pass deepest into the potential well of either the MW or M31 are Leo~I (MW) and IC~10 (M31).\footnote{For Leo I, \citet{Sohn_2013} has illustrated the effects of using a 1:1, 1:2, and 2:1 mass ratio for the MW and M31, concluding that all resulting mass ratios yield a recent pericentric passage around the MW. The most notable difference in the results as the mass of the MW increases is that a second pericentric passage becomes possible within the orbital uncertainties, but it is statistically rare. We expect that IC~10 would also be minimally affected given the similarity between the Leo~I-MW orbit and the IC~10-M31 orbit.} 

Corresponding disk and bulge masses and their associated scale radii are listed in Table 2 of \citet{patel17a}. These values were determined by fitting for the parameters that yield the closest match to the observed rotation curves for the MW \citep{mcmillan11} and M31 \citep{corbelli10}. Both the MW and M31's dark matter halos are adiabatically contracted using the \texttt{CONTRA} code \citep{contra}, and the halos are truncated at their respective virial radii.

A dynamical friction (DF) prescription following the Chandrasekhar formula \citep{chandrasekhar} is implemented to account for the DF owing to both the MW and M31. The strength of DF is determined by the Coulomb logarithm (ln$\Lambda$). For the LMC and M33, the Coulomb logarithm is the same as in \citet{patel17a}, where the parametrization provided for a 1:10 mass ratio in the Appendix of \citet{vdm12ii} is adopted. For DF between the MW and M31, DF is prescribed using the equal mass Coulomb logarithm from the Appendix of \citet{vdm12ii}. For the target galaxy, or the least massive galaxy in a 5-body system, the Coulomb logarithm follows that of \citet{hashimoto03} where the DF term varies as a function of the relative distance between the target galaxy and the MW/M31 or the LMC.

\begin{figure*}[t]
    \centering
    \includegraphics[scale=0.5
]{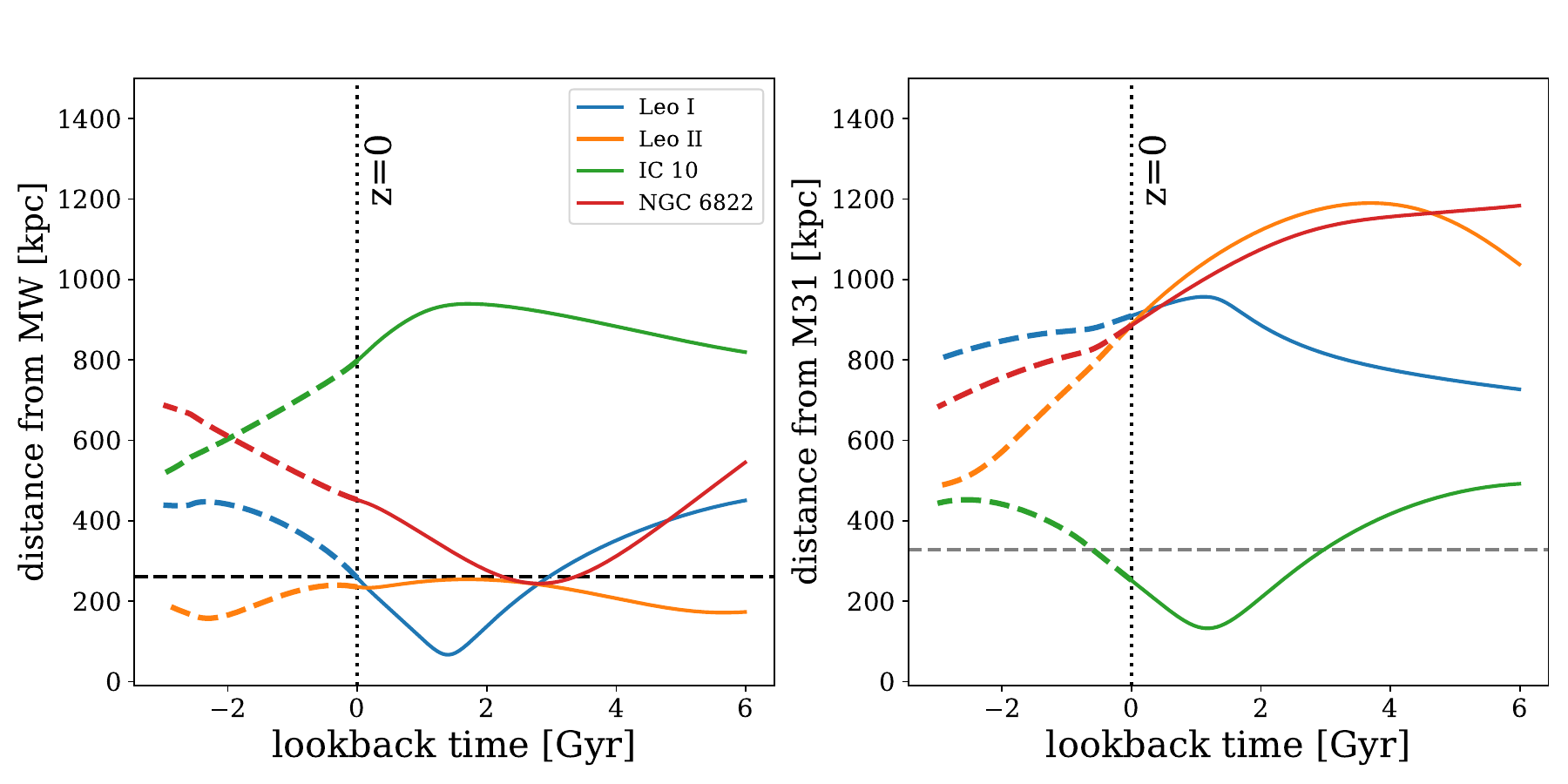}
    \includegraphics[scale=0.5]{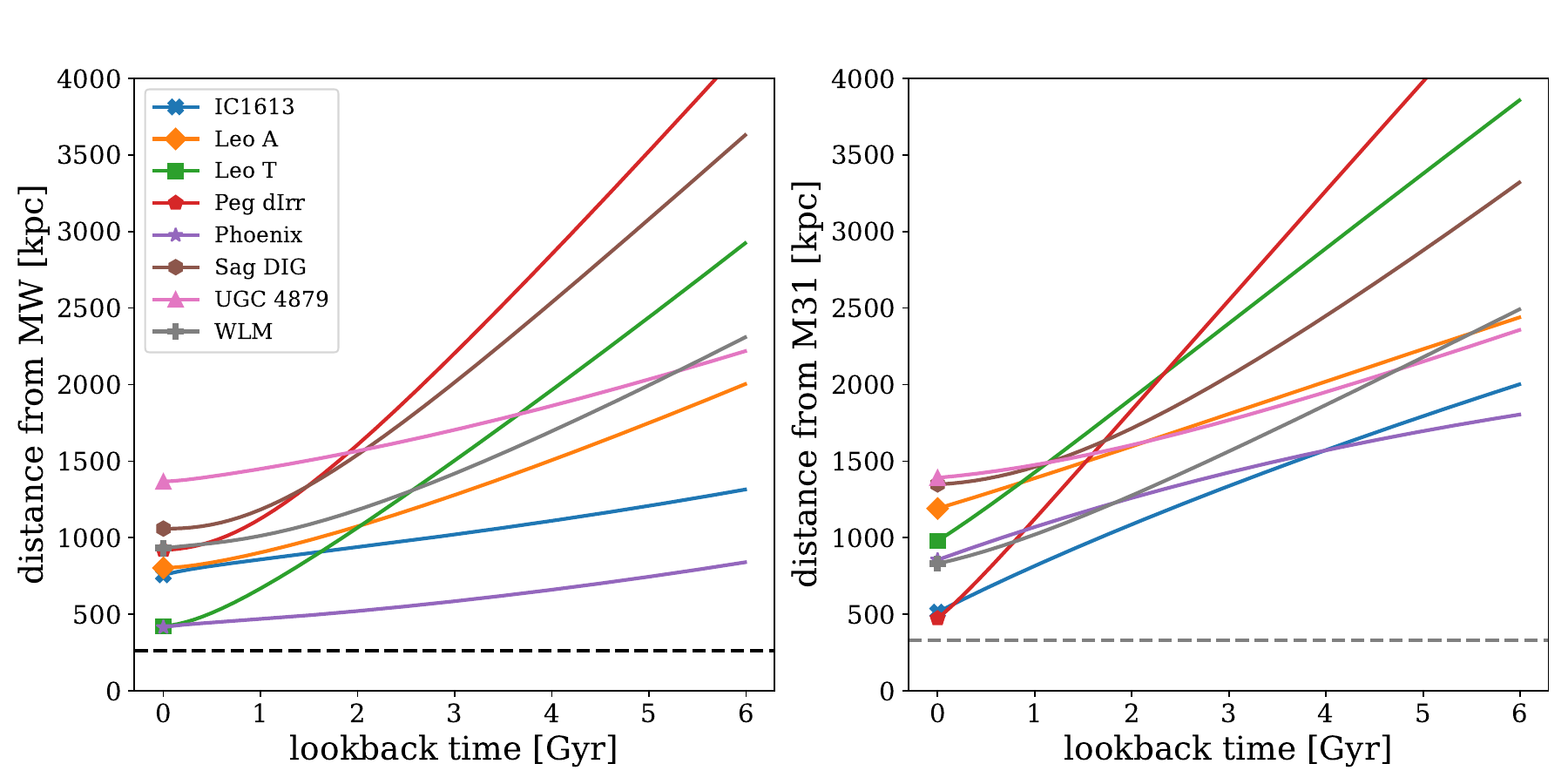}
    \caption{\textbf{Top:} The direct orbital histories for four galaxies in our sample spanning the range of 2 Gyr into the future and going backward to 6 Gyr in the past. Present day is denoted with a dotted black line. The left panel shows orbits relative to the MW, while the right panel shows orbits with respect to M31. The black and gray dashed lines indicate the virial radii of the MW and M31, respectively. The four galaxies pass within the virial radius of either the MW (for the case of Leo~I, Leo~II, and NGC~6822) or M31 (for IC~10) during the 8 Gyr time period plotted, suggesting that these galaxies may be considered interacting or backsplash galaxies. \textbf{Bottom:} Similar as the top panels for the remaining eight galaxies in our sample, but showing only the past 6 Gyr and with a more extended vertical axis. These galaxies are all on first infall into the MW and M31 as the have either never crossed the virial radius of the MW or M31 or have only crossed it very recently. }
    \label{fig:orbits}
\end{figure*}

The LMC and M33 are modeled as single component potentials consisting of Plummer spheres with masses of $1.8\times 10^{11}\, M_{\odot}$ and $2.5\times 10^{11}\, M_{\odot}$, respectively. These masses are identical to those used in \citet{patel17a}, thus we refer readers to that work for details on how these masses were chosen. All galaxies in our sample with the exception of IC~10, Leo~I and Leo~II are considered point masses of $10^{10}\, M_{\odot}$. We have adopted the following scale lengths for the three mentioned systems,  computed by fitting Plummer \citep{plummer11} spheres to their dynamical masses: IC~10 (2.1 kpc), Leo~I (2.3 kpc), Leo~II (2.1 kpc). These non-zero halo scale lengths are important for approximating the strength of dynamical friction owing to the MW or M31 via the Coulomb logarithm. 

\begin{figure*}[h]
    \centering
    \includegraphics[scale=0.5]{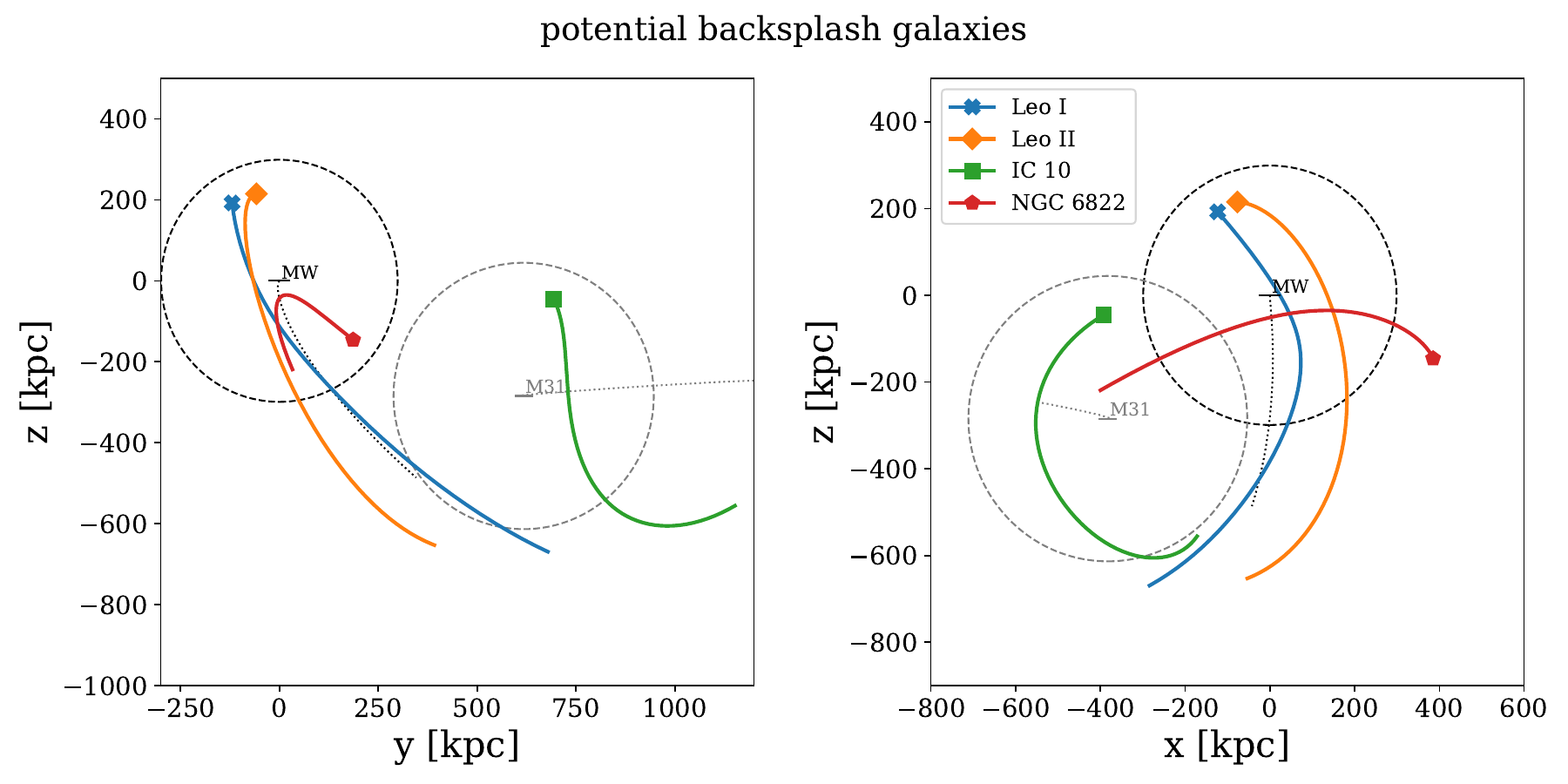}
    \includegraphics[scale=0.5]{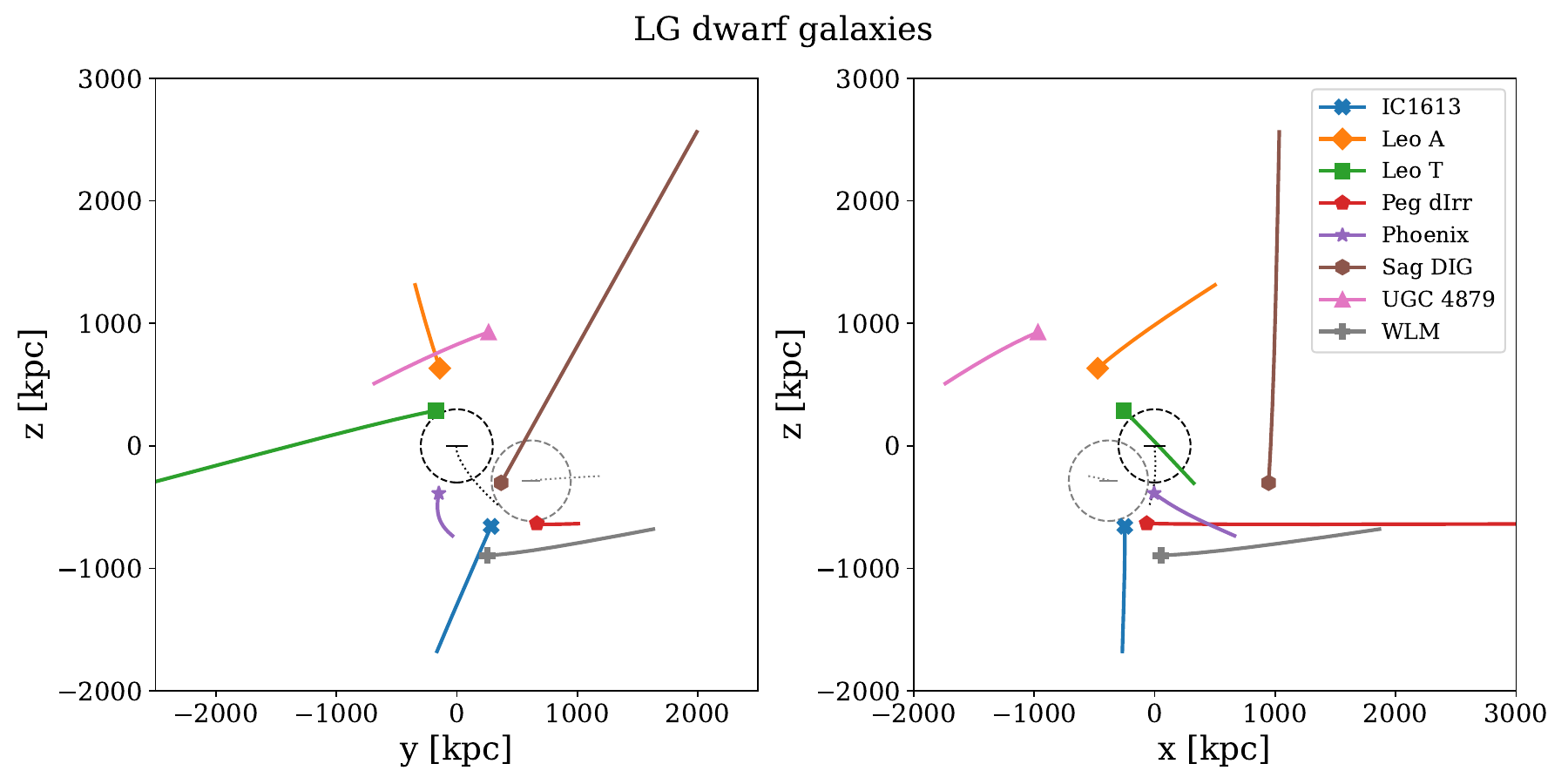}
    \caption{A spatial projection of the direct orbital histories illustrated in Figure \ref{fig:orbits} in Galactocentric coordinates. The top panels show likely backsplash galaxies, while the bottom panels show likely first-infall galaxies, this is the same as in Figure \ref{fig:orbits}. In all panels, the virial extent of the MW (black) and M31 (gray) are indicated with dotted circles. Their present day positions are marked with horizontal dashes. The dotted gray and black lines indicate the motion of the MW and M31 over the last 6 Gyr. For all galaxies in our sample, marker symbols also indicate their present day location. The top panel shows that Leo~I, Leo~II, and NGC~6822 have crossed through the virial radius of the MW. Similarly, IC~10 has passed through the virial radius of M31 in the last 6 Gyr. The bottom panels show the first infall galaxies in our sample. Many of them were hundreds to thousands of kiloparsecs from the MW or M31 6 Gyr ago, having only recently entered the LG and/or the virial extents of the MW/M31.}
    \label{fig:xy_yz_orbits}
\end{figure*}

Though all five gravitational potentials are simultaneously integrated backward in time, the least massive galaxy (i.e., the target galaxy from the sample considered in this paper) does not impart an acceleration on the other four bodies. Thus, each target galaxy feels the influence of four other bodies but the MW, M31, LMC, and M33 only experience the gravitational influence of each other (i.e, a true 4-body encounter). Orbits are integrated for a period of 6 Gyr into the past using a Leapfrog algorithm as in our previous work \citep[e.g.,][]{patel17a, patel20, sohn20}, with the exception of Leo~I, Leo~II, IC~10, and NGC~6822 which are also integrated 2 Gyr into the future for a total of 8 Gyr. These orbits will be referred to as \emph{direct} orbital histories as they use the 3D position and velocity vectors directly from the transformation of LOS velocity, distance and PMs to 6D Galactocentric coordinates as initial conditions (see Table \ref{tab:Pos_vel}). Uncertainties on orbital histories will be discussed in Section \ref{subsec:orbits}.

We only integrate backward for a period of 6 Gyr for several reasons. First, our rigid potentials do not account for mass evolution or mass loss due to tides. Regarding the mass evolution of the MW, for example, \citet{santistevan20} find that MW-like galaxies have assembled about 80\% of their mass by 6 Gyr ago\footnote{In principle, virial radii are also about 80\% of their present day values at 6 Gyr ago, however, halo concentration is more likely to change over the at a fixed halo mass, even in the time frame consider \citep{diemer15}. As we use rigid halos, concentration is held fixed in our models.}. Second, going further back would require an approximation for the MW and M31's mass evolution to recover accurate orbits, and furthermore, orbits become less certain moving further back in time. Finally, we do not opt for a shorter time period as many of the satellites in the Local Group take between $\sim$ 2-5 Gyrs to complete one orbit about their host galaxy, thus a shorter time frame would not provide as much insight on the evolutionary history of galaxies in our sample. For more details on how orbit recovery compares to orbits in cosmological simulations, see the Appendix of \citet{patel17a}, \citet{dsouza22,santistevan23}.

\begin{deluxetable*}{lcccccc}[h]
\tablenum{6}
\tablecaption{Orbital Parameters Relative to the Milky Way\label{tab:orbit_params_mw}}
\tablewidth{0pt}
\tablehead{
\colhead{Dwarf}  &  \colhead{$\rm f_{peri}$} & \colhead{$\rm t_{peri}$} & \colhead{$\rm r_{peri}$} &  \colhead{$\rm f_{apo}$} & \colhead{$\rm t_{apo}$} & \colhead{$\rm r_{apo}$} \\
\multicolumn1c{} & \colhead{(\%)} &  \colhead{(Gyr)} & \colhead{(kpc)} &  \multicolumn1c{(\%)} & \multicolumn1c{(Gyr)}  & \multicolumn1c{(kpc)} 
}
\startdata
IC~10 & 22.7 & $\cdots$ [3.92, 5.68] & $\cdots$ [676, 900] & 83.4 & 1.72 [1.13, 3.3] & 939 [913, 1007] \\
Leo~I & 100.0 & 1.4 [1.32, 1.53] & 66 [64, 86] & 7.4 & $\cdots$ [5.3, 5.91] & $\cdots$ [331, 370] \\
Leo~II & 100.0 & 0.17 [0.14, 0.18] & 233 [220, 247] & 80.6 & 1.7 [1.22, 3.13] & 254 [236, 290] \\
NGC~6822 & 100.0 & 2.81 [0.78, 2.86] & 243 [170, 424] & 0.2 & $\cdots$ [2.85, 5.17] & $\cdots$ [290, 391] \\ \hline
IC~1613 & 0.0 & $\cdots$ [0.0, 0.0] & $\cdots$ [0, 0] & 0.8 & $\cdots$ [4.49, 5.45] & $\cdots$ [956, 1032] \\
Leo A & 0.0 & $\cdots$ [0.0, 0.0] & $\cdots$ [0, 0] & 7.6 & $\cdots$ [5.21, 5.88] & $\cdots$ [1065, 1191] \\
Leo T & 0.0 & $\cdots$ [0.0, 0.0] & $\cdots$ [0, 0] & 0.0 & $\cdots$ [0.0, 0.0] & $\cdots$ [0, 0] \\
Peg dIrr & 16.0 & $\cdots$ [0.01, 1.61] & $\cdots$ [865, 945] & 6.1 & $\cdots$ [0.11, 0.24] & $\cdots$ [884, 945] \\
Phoenix & 0.5 & $\cdots$ [3.83, 5.55] & $\cdots$ [293, 459] & 11.2 & $\cdots$ [1.62, 4.08] & $\cdots$ [446, 505] \\
Sag DIG & 100.0 & 0.05 [0.04, 0.11] & 1059 [978, 1143] & 0.0 & $\cdots$ [0.0, 0.0] & $\cdots$ [0, 0] \\
UGC 4879 & 3.3 & $\cdots$ [0.01, 0.04] & $\cdots$ [1343, 1392] & 0.6 & $\cdots$ [4.77, 5.82] & $\cdots$ [1594, 1715] \\
WLM & 4.4 & $\cdots$ [1.72, 5.17] & $\cdots$ [831, 955] & 7.3 & $\cdots$ [0.31, 0.6] & $\cdots$ [915, 976] \\
\enddata
\tablecomments{Orbital parameters calculated relative to the MW. Parameters are taken from the direct orbital histories shown in Figure \ref{fig:orbits}, while values in brackets quote the [15.9, 84.1] percentiles around the median calculated from 1,000 Monte-Carlo orbits that propagate the measured uncertainties on distances, LOS velocities, and PMs. Dots signify that a quantity is not recovered in the individual direct orbital history (e.g., there was no pericenter or apocenter passage over the 6 Gyr orbital integration period). Values of [0,0] imply that the quantity is not recovered for any of the Monte-Carlo calculated orbits. Columns 2 and 5 indicate the percentage of the 1,000 orbits that passed through peri-/apocenter in the last 6 Gyr time period. Note that to be considered a satellite, a galaxy must reside inside the virial radius of its host galaxy today (261 kpc for the MW and 329 kpc for M31).} 
\end{deluxetable*}

\begin{deluxetable*}{lcccccc}[h]
\tablenum{7}
\tablecaption{Orbital Parameters Relative to Andromeda\label{tab:orbit_params_m31}}
\tablewidth{0pt}
\tablehead{
\colhead{Dwarf}  &  \colhead{$\rm f_{peri}$} & \colhead{$\rm t_{peri}$} & \colhead{$\rm r_{peri}$} &  \colhead{$\rm f_{apo}$} & \colhead{$\rm t_{apo}$} & \colhead{$\rm r_{apo}$} \\
\multicolumn1c{} & \colhead{(\%)} &  \colhead{(Gyr)} & \colhead{(kpc)} &  \multicolumn1c{(\%)} & \multicolumn1c{(Gyr)}  & \multicolumn1c{(kpc)} 
}
\startdata
IC~10 & 100.0 & 1.18 [0.82, 1.35] & 132 [80, 184] & 36.4 & $\cdots$ [4.35, 5.6] & $\cdots$ [320, 442] \\
Leo~I & 38.8 & $\cdots$ [2.78, 4.93] & $\cdots$ [816, 938] & 100.0 & 1.12 [1.05, 1.31] & 957 [941, 995] \\
Leo~II & 2.8 & $\cdots$ [5.0, 5.89] & $\cdots$ [949, 1240] & 82.5 & 3.69 [3.12, 4.82] & 1190 [1143, 1287] \\
NGC~6822 & 18.8 & $\cdots$ [2.03, 4.55] & $\cdots$ [633, 935] & 21.5 & $\cdots$ [1.44, 3.68] & $\cdots$ [913, 1142] \\ \hline
IC~1613 & 0.0 & $\cdots$ [0.0, 0.0] & $\cdots$ [0, 0] & 0.0 & $\cdots$ [0.0, 0.0] & $\cdots$ [0, 0] \\
Leo A & 0.2 & $\cdots$ [5.89, 5.95] & $\cdots$ [1070, 1098] & 58.3 & $\cdots$ [1.42, 5.07] & $\cdots$ [1113, 1249] \\
Leo T & 7.3 & $\cdots$ [0.05, 0.91] & $\cdots$ [885, 995] & 1.0 & $\cdots$ [0.9, 5.14] & $\cdots$ [986, 1204] \\
Peg dIrr & 19.5 & $\cdots$ [0.06, 0.82] & $\cdots$ [270, 473] & 0.5 & $\cdots$ [0.92, 5.56] & $\cdots$ [490, 591] \\
Phoenix & 0.2 & $\cdots$ [5.59, 5.89] & $\cdots$ [1244, 1251] & 8.0 & $\cdots$ [4.27, 5.76] & $\cdots$ [1103, 1357] \\
SAG DIG & 0.0 & $\cdots$ [0.0, 0.0] & $\cdots$ [0, 0] & 0.0 & $\cdots$ [0.0, 0.0] & $\cdots$ [0, 0] \\
UGC 4879 & 39.4 & $\cdots$ [0.16, 1.66] & $\cdots$ [1187, 1381] & 0.0 & $\cdots$ [0.0, 0.0] & $\cdots$ [0, 0] \\
WLM & 28.0 & $\cdots$ [0.12, 1.73] & $\cdots$ [613, 832] & 2.5 & $\cdots$ [1.08, 4.33] & $\cdots$ [832, 955] \\
\enddata
\tablecomments{Same as Table \ref{tab:orbit_params_mw} but all quantities are computed with respect to M31. }
\end{deluxetable*}

\subsection{Orbital Histories and Analysis}
\label{subsec:orbits}

The direct orbital histories for all galaxies in our sample are illustrated in Figure \ref{fig:orbits}. The top row shows the orbits of Leo~I, Leo~II, IC~10, and NGC~6822 where the left panel shows the orbits relative to the MW and the right panel shows the orbits of these galaxies relative to M31. 
Present day ($z=0$) is indicated with a vertical dotted line. The virial radii of the MW or M31 are indicated with gray and black dashed lines, respectively. The orbits of Leo~I, Leo~II, and NGC~6822 pass within the virial radius of the MW either in the past or future. Similarly, IC~10 undergoes pericentric passage with M31 about 1 Gyr ago. The orbital parameters associated with the direct orbital histories are listed in Tables \ref{tab:orbit_params_mw} and \ref{tab:orbit_params_m31}, measured relative to the MW and M31, respectively.

The bottom row of Figure \ref{fig:orbits} shows the orbits of all remaining galaxies in our sample, again with respect to the MW on the left and M31 on the right. The direct orbits of these galaxies suggest they are first infall into the halos of the MW, M31, or the LG as a whole, however, it should be noted that the PM uncertainties of many of these galaxies lead to large 3D velocity errors (see Table \ref{tab:Pos_vel}) and can therefore bias orbits towards first infall since orbital uncertainties grow moving backwards in time. While these galaxies are at their closest distances at present day and were between $\sim$0.5-4 Mpc from the MW or $\sim$1.5-6 Mpc from M31 at 6 Gyr ago, future PM improvements will yield more refined orbital histories that may or may not find first infall as the most statistically common orbital history. The nature of these galaxies versus those that pass within the virial radii of the MW or M31 will be discussed further in Section \ref{sec:disc}. 

Figure \ref{fig:xy_yz_orbits} shows cross sections of the direct orbital histories shown in Figure \ref{fig:orbits}. In Figure \ref{fig:xy_yz_orbits}, orbits are plotted for the last 6 Gyr for all galaxies in our sample (i.e., no future orbits are included). The top panel shows the orbital histories of Leo~I, Leo~II, IC~10, and NGC~6822. The symbols represent their locations today and the lines trace their orbits backwards in time. The virial radii of the MW and M31 are indicated with circles. The orbital histories of the MW and M31 are also shown, as these galaxies are allowed to move in response to the gravitational influence of other galaxies in our orbital models. These cross sections clearly indicate that Leo~I, Leo~II, and IC~10 have passed into the virial radii of the MW or M31 one time in the last 6 Gyr. However, the history of NGC~6822 is a bit more complex at first glance. It appears to have crossed into and out of the halo of the MW with an origin in the halo of M31 in projection. However, combined with the view of its orbital history as illustrated in Figure \ref{fig:orbits}, NGC~6822 has never been closer than $\sim$ 800 kpc from M31. Additionally it only briefly passed into the edge of the MW's halo (defined by our choice of virial radius). We will further discuss the unusual orbit of NGC~6822 and its connection to its recent SFH in Section~\ref{sec:disc}.

The bottom panels of Figure \ref{fig:xy_yz_orbits} show the remaining eight galaxies in our sample in projection about the MW and M31. As seen in Figure \ref{fig:orbits}, these galaxies were located far from the MW or M31 6 Gyr ago and have only recently entered the vicinity of the LG. Leo T and Phoenix come closest to the MW at present day, while IC~1613 and Peg dIrr come the closest to M31 at present day. Regardless of their proximity today, the orbits shown in Figures \ref{fig:orbits} and \ref{fig:xy_yz_orbits} do not suggest that these galaxies can be considered satellites of the MW or M31 at present with the possible exception of Peg dIrr (see below). We note that the orbits in Figures \ref{fig:orbits} and \ref{fig:xy_yz_orbits} do not yet account for the measurement uncertainties corresponding to each galaxy's PM, LOS velocity, and distance. To understand how statistically likely the orbital histories illustrated in these figures are, we must account for these uncertainties and the range of orbital histories that can result from them. 

As in previous work \citep[e.g.,][]{patel17a, patel20, sohn20}, to investigate the uncertainties on the orbits of galaxies in our sample, we run 1,000 orbits for each galaxy using position and velocity vectors that are drawn in a Monte Carlo fashion from the measurement uncertainties on the distance, LOS velocity, and PM of each galaxy. From this ``error propagation" we determine the resulting uncertainty on the orbital parameters for each galaxy, including the pericenter/apocenter distance and timing. 
Tables~\ref{tab:orbit_params_mw} and~\ref{tab:orbit_params_m31} list the results for the direct orbits, together with the Monte-Carlo uncertainties corresponding to the [15.9. 84.1] percentiles around the median for each quantity. We also list what fraction of Monte-Carlo orbits have completed a pericenter and/or apocenter. 

\begin{figure*}[t]
\centering
    \includegraphics[scale=0.55]{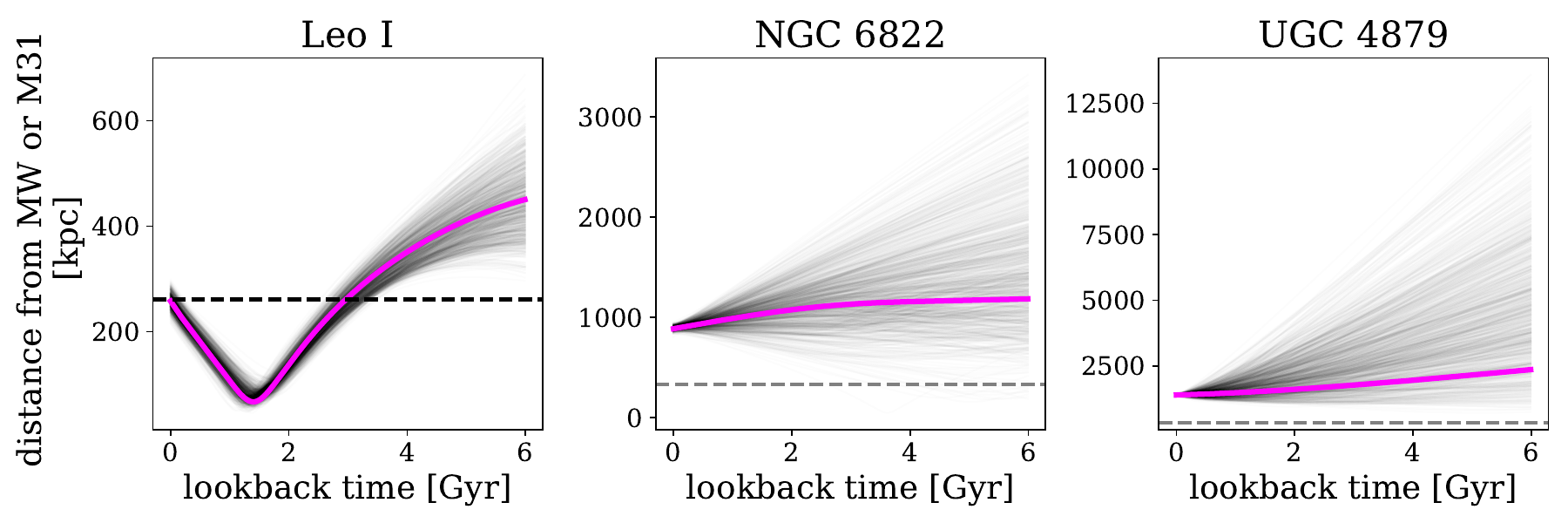}
    \caption{An illustration of the range of uncertainties in orbital histories for galaxies with small (left), intermediate (middle), and large PM uncertainties (right). Leo I is plotted with respect to the MW, while NGC~6822 and UGC~4879 are plotted with respect to M31. Each panel also includes the virial radius of the MW or M31 in a black or dashed gray line, respectively. The magenta curve is the direct orbital history for each galaxy (i.e., the same as the direct orbits in Figure \ref{fig:orbits}). The gray lines represent 1,000 orbits spanning the uncertainty in the 6D phase space coordinates in Table \ref{tab:Pos_vel}. As expected, orbital uncertainties increase as a function of increasing distance.}
    \label{fig:errors}
\end{figure*}

Figure \ref{fig:errors} shows an illustration of orbital uncertainties with low (Leo~I), intermediate (NGC~6822), and high (UGC 4879) PM uncertainties. All orbits are plotted with respect to the MW. The magenta lines indicate the same direct orbital histories as in Figure \ref{fig:orbits}, while the gray lines indicate the 1,000 orbital histories resulting from the error propagation of the measured uncertainties. For Leo~I, it is clear that all orbital histories agree and only show a significant scatter at around $\sim$6 Gyr. For NGC~6822, the depth at pericenter around $\sim$3 Gyr varies across the 1,000 orbits from $\sim$170-420 kpc (see Table \ref{tab:orbit_params_mw}). For UGC 4879, all orbits indicate a first approach towards the MW but orbits can vary by hundreds of kiloparsecs even in the recent past ($<$ 2 Gyr). 

Table \ref{tab:orbit_params_mw} shows that Leo~I, Leo~II, and NGC~6822 complete a pericentric passage around the MW in 100\% of orbits. The typical distance at pericenter for these three galaxies is always within the virial radius of the MW \citep[see][261 kpc]{patel17a}. All of the Monte-Carlo orbits for Sag DIG ($f_{peri}=$100\%) complete a passage around the MW, however, the distance at pericenter is very large ($\sim$1 Mpc), thus this galaxy is a LG dwarf rather than being associated with either the MW or M31, where a satellite of the MW or M31 is simply defined as a galaxy presently within the virial radius of the MW or M31. Some other galaxies in the sample have non-zero $f_{peri}$ percentage values, however, they too orbit around the MW at very large distances. None of the galaxies in the sample have scenarios where 100\% of orbits recover both a pericenter and apocenter in the last 6 Gyr. The galaxy that reaches closest to this is Leo~II for which 100\% of orbits pass through pericenter and 80.6\% of orbits pass through an apocenter approximately within the MW's virial radius. 

Table \ref{tab:orbit_params_m31} indicates that only IC~10 shows any kind of long-term statistically significant association with M31. All 100\% of IC~10's Monte-Carlo orbits pass through pericenter well within M31's virial radius (329 kpc). Peg dIrr was within the M31 virial for 19.5\% of orbits, so even if it is on first infall, it may be a (just arriving) M31 satellite.

\subsection{Comparison to previous orbits}

Some of the dwarfs in our sample have reported orbits calculated using PMs in the literature. In this section, we compare these to our newly derived orbits. \\

\noindent \textbf{IC~10:} The orbit of IC~10 has been studied in detail in \citet{Nidever_2013} and \citet{Namumba_2019}. \citet{Nidever_2013} use the PM measurement from \citet{Brunthaler_2007} to calculate the possible orbital history of IC~10 around M31, adopting a mass of $1.4\times10^{12}\, M_{\odot}$ for M31, though recent results suggest the mass of M31 might be twice as massive \citep{patel23}. The most recent pericentric passage in the resulting orbit occurs at $1.88^{+0.34}_{-0.04}$ Gyr ago at a distance of $r_{\rm peri}=82^{+94}_{-26}$ kpc. This orbital history is approximately in agreement with the results reported in Table \ref{tab:orbit_params_m31} which indicate a pericentric passage at $1.18^{+0.17}_{-0.36}$ Gyr ago at a distance of $r_{\rm peri}=132^{+52}_{-52}$ kpc. Both our orbital histories and that reported in \citet{Nidever_2013} would designate it as a galaxy with a recent pericentric passage based on its orbital properties in the last few billion years.

\cite{Namumba_2019} carry out N-body simulations of the M31-IC~10 system to determine whether tidal interactions with M31 could have formed extensions of HI gas observed around IC~10. They explore a set of possible orbits within the measurement uncertainties of the \citet{Brunthaler_2007} PMs and find that the HI features observed around IC~10 are unlikely to be due to a recent tidal interaction. However, they do not report orbital parameters resulting from their suite of simulations, thus a direct comparison cannot be made to our results. \\

\noindent \textbf{Leo~I:}
Orbits for Leo~I were calculated in \cite{Sohn_2013}. Their orbit shows substantial similarities to ours, agreeing that Leo~I is a galaxy that is currently receding from the Milky Way following a recent close approach. However, there are small differences in the derived orbits due to the differences in PMs derived between this work and \cite{Sohn_2013} (see Figure \ref{figure:LeoI}) and advances in orbit modeling over the previous 10 years. Differences include the timing of the most recent pericenteric passage. Here we report $t_{peri}=1.40^{+0.13}_{-0.08}$ Gyr and \citet{Sohn_2013} concludes $t_{peri}=1.05\pm0.09$ Gyr. The distance of Leo~I relative to the MW at pericenter also differs. Here, we find $r_{peri}=66^{+20}_{-2}$ kpc, whereas \citet{Sohn_2013} finds a marginally larger distance of $r_{peri}=91\pm36$. 
These differences are important for the comparison between orbital and SFHs (see \S \ref{subsec:SFH}) as they change the timing of key events such as Leo~I entering the virial radius of the MW for the first time. \\

\noindent \textbf{Leo~II:} Orbits for Leo~II were independently calculated using PMs measured with \textit{Gaia} DR3 data by \citet{Li_2021, Battaglia_2022, Pace_2022}. \citet{Li_2021} adopt an NFW halo for the MW with a mass of $8.1 \times 10^{11} \, M_{\odot}$ and neglect the influence of the LMC on the orbit of Leo~II. They conclude that Leo~II's most recent pericentric passage about the MW occured at $r_{\rm peri}=114^{+104}_{-75}$ kpc with $r_{\rm apo}=254^{+146}_{-16}$. The latter value for the distance at apocenter is in excellent agreement with the results we report in Table \ref{tab:orbit_params_mw}, however, we find a pericentric distance significantly larger than \cite{Li_2021} at $r_{\rm peri}=233^{+14}_{-13}$ kpc. Our orbital results suggest an approximately circular orbit for Leo~II, whereas \citet{Li_2021} find an orbit halfway between circular and elliptical with an eccentricity of $e=0.47$.

\citet{Battaglia_2022} consider both a light and heavy MW, where their light MW is most consistent with our model. They also include the perturbations induced by the LMC on the MW. In this perturbed ``light MW" model, Leo~II reaches pericenter at $\sim$ 3 Gyr ago at a distance of $\sim$65 kpc and apocenter at $\sim$5.5 Gyr ago at a distance of $\sim$ 205 kpc. These values come from orbits integrated without including the measurement uncertainties on distance, PMs, and LOS velocity (similar to our ``direct" orbital histories). \citet{Battaglia_2022}'s pericentric distance is significantly smaller than our conclusions listed in Table \ref{tab:orbit_params_mw}, and even smaller than those reported in \citet{Li_2021}. 

Finally, \citet{Pace_2022} consider a MW mass of $1.3\times10^{12} \,M_{\odot}$ and the effect of the LMC ($1.38\times 10^{11} \, M_{\odot}$) to integrate the orbits of MW satellites. For Leo~II, they find that the most recent pericentric passage about the MW occurred at $r_{\rm peri}=61.4^{+62.3}_{-34.7}$ kpc with $r_{\rm apo}=230^{+17.6}_{-17.1}$. When the LMC is not included, their distance at pericenter decreases by approximately 20\% and the distance at apocenter increases by a few percent. Overall, these comparisons show that our distance at pericenter for Leo~II is three to four times higher than that found in the three studies discussed above. 

These differences are likely attributed to the varying MW and LMC mass models (if an LMC is included), as well as any differences in the measured PMs. To test this, we use the PMs reported in \citet{Li_2021, Battaglia_2022, Pace_2022} with our orbital setup to see how much the PMs themselves contribute to differences in the resulting orbital histories. At face value, the PMs we report are most similar to those of \citet{Li_2021}, therefore the orbit we compute using their PMs is nearly identical to the Leo~II orbit we propose for the last 6 Gyr in Figures \ref{fig:orbits} and \ref{fig:xy_yz_orbits}. The orbits resulting from the PMs of \citet{Battaglia_2022} and \citet{Pace_2022} are very similar to each other (as are their PMs). These orbits indicate that Leo~II has a pericentric passage between 3-4 Gyr ago at distances of $\sim$100 kpc. The most significant difference in the 3D velocity vectors between these four sets of results is that the x-component of velocity can differ by up to $\sim$ 40 km s$^{-1}$, as is the case for our velocity vector compared to \citet{Pace_2022}. Further improvements in the PM measurements for Leo~II will be necessary to definitely determine its orbital history. \\

\noindent \textbf{Leo T:} A series of potential orbits for Leo T are calculated in \cite{Blana2020}. These explore a number of possible orbits using the position, LOS velocity, and gas properties of Leo T to predict possible PMs and explore which orbital histories fit with different PM predictions. 

Our orbital calculation corresponds best to what \cite{Blana2020} refers to as ``case 3a", a static MW potential of $1\times 10^{12}\, M_{\odot}$ with the inclusion of M31. However, this orbital calculation does not take into account the impact of the LMC, which has been shown to have a significant impact on the MW and its satellites \citep[e.g.,][]{patel20}. Despite this, we find agreement with the results from \cite{Blana2020} which show that Leo T would be on first infall with the PM measured in this work in all cases examined in \cite{Blana2020}. \\

\noindent \textbf{Phoenix:} An orbit for Phoenix was calculated by \cite{Battaglia_2022} using PM results from \textit{Gaia} DR3 data. This showed a likely first-infall origin for Phoenix but could not completely exclude a backsplash origin. Our smaller PM uncertainties allow us clearly show that Phoenix is on first-infall rather than being a backsplash galaxy. \\




\noindent \textbf{NGC~6822:} \cite{Battaglia_2022} also calculated an orbit for NGC~6822 using the PM results from \textit{Gaia} DR3 data. This has excellent agreement with our derived orbit and shows that NGC~6822 has had a pericentric passage around the MW $\sim$3 Gyr ago but the large uncertainty in its orbit does not allow us to make any strong claim about the depth of this passage within the MW potential (see Figure~\ref{fig:errors}). 

\section{Discussion} \label{sec:disc}


\begin{deluxetable*}{c|c|c|c}
\tablenum{8}
\tablecaption{Probability of dwarfs being a backsplash galaxy \label{tab:splash}}
\tablewidth{0pt}
\tablehead{
\colhead{Dwarf}  &  \colhead{Probability} &  \colhead{Probability of} & \colhead{Probability of}  \\
\colhead{Name}  &  \colhead{Backsplash from MW} & \colhead{Backsplash from M31} & \colhead{Backsplash Total} \\
\multicolumn1c{} & \colhead{This Work} & \colhead{This Work} &  \colhead{\cite{Buck_2019}}
}
\startdata
IC~1613 & 0\% & 0\% & 61\% \\
Leo A & 0\% & 0\% & -  \\
Leo T & 0\% & 0\% & 51\% \\
NGC~6822 & 27.6\% & 0.9\% & 61\% \\
Peg dIrr & 0\% & 5.2\% & 79\%   \\
Phoenix & 0.1\% & 0\% & 43\% \\
Sag DIG & 0\% & 0\% & - \\
UGC 4879 & 0\% & 0\% & - \\
WLM & 0\% & 0.1\% & 0\% \\
\enddata
\tablecomments{Probabilities of dwarf galaxies in our sample being backsplash galaxies compared to those computed using radial velocity and position in \cite{Buck_2019} where available. This work defines a backsplash galaxy as one that is currently outside of the virial radius of the MW (M31) and has previously undergone a pericentric passage within the virial radius of the MW (M31). Therefore, percentages represent the fraction relative to all 1,000 orbits computed for each satellite where a pericenter at a distance less than the virial radius was recovered. The probability of a dwarf in this table being on first infall is 100\%-(combined M31 \& MW backsplash probability).}
\end{deluxetable*}

\subsection{Backsplash Galaxies} \label{subsec:backsplash}

An important way to explore this sample of dwarf galaxies is to examine their interaction history, i.e. if those galaxies have interacted previously with a massive host and, if so, when and for how long. This is particularly important for those galaxies outside of the virial radius of the MW or M31 at the present time, to see if they are on first infall (i.e., arriving at their closest position relative to the MW or M31 today without having previously passed around the primary galaxy) or if they have previously had a pericentric passage within the virial radius of the MW or M31 (i.e. are backsplash galaxies). 

Previous work has examined this problem via radial velocity and positions \citep[][]{Buck_2019}. We find that the introduction of PMs, and therefore full 6D position-velocity information, makes determinations easier as we examine a much more limited area of parameter space. Table~\ref{tab:splash} shows for those galaxies in our sample currently outside the virial radius of the MW or M31, the fraction relative to all 1,000 orbits computed for each satellite where a pericenter at a distance less than the virial radius was recovered. 
Probabilities from \citet[][]{Buck_2019} are listed for reference. For most galaxies, we find substantially more certain results, with almost all galaxies in our sample outside the virial radius being either certainly 
or almost-certainly on first infall (except possibly NGC~6822, as discussed below). This determination is in favor of galaxies being on first infall, even in cases where the probabilities seemed to favor a backsplash-galaxy interpretation in previous work. 



NGC~6822 is a more uncertain case. This galaxy has a pericentric passage about the Milky Way in all orbits (see Table \ref{tab:orbit_params_mw}) and this pericenter is within the MW virial radius in 27.6\% of orbits (see Table \ref{tab:splash}). Therefore, it has a non-negligible probability of being classified as a backsplash galaxy. However, the direct orbit pericenter distance is $r_{\rm peri} = 243$ kpc, with only a small fraction of Monte-Carlo orbits penetrating significantly into the MW dark matter halo (see Table~\ref{tab:orbit_params_mw} and Figure~\ref{fig:errors}). Any passage through the MW's virial radius was likely fleeting and would have had very little effect on the dwarf. We conclude it may have undergone a brief 'flyby encounter', but it is not a  'true' backsplash galaxy. 

The remaining galaxies in our sample not listed in Table~\ref{tab:splash} (namely, IC~10, Leo~I and Leo~II) are those that are inside the virial radius of either the MW or M31 at the present time. By definition, these galaxies are still undergoing an interaction with a massive host and are therefore not backsplash galaxies at the present time. However, if we examine the future trajectory of these galaxies (see Figure \ref{fig:orbits}) we find that IC~10 and Leo~I are destined to become backsplash galaxies within the next 1~Gyr. 

\subsection{Star formation Histories} \label{subsec:SFH}

By comparing the galaxies' orbital histories presented in Section \ref{sec:orbits} to their measured SFHs from the literature, we can look for correlations that may provide more insight into how and when these galaxies formed their stars or ceased to form stars. This may provide clues about how the environment of the dwarf galaxies in our sample affects their star formation.  

Detailed comparisons are stymied not only by uncertainties in the orbital calculations (quantified through the Monte-Carlo approach discussed in Section~\ref{sec:orbits}), but also by uncertainties in the SFHs. 
The SFH uncertainties are smallest for those galaxies where the available CMDs reach the old Main Sequence Turnoff (oMSTO). This is the case for our sample galaxies IC 1613, Leo I, Leo II, Leo A, Leo T, Phoenix, and WLM (see references provided in subsections below). Shallower CMDs (as available for IC10, NGC 6822, Pegasus, Sag DIG, and UGC 4879) are generally less reliable, especially for constraining the exact amount of oldest star formation. But for correlating orbital characteristics with SFHs in the last few Gyr, deep CMDs that reach the oMSTO may be less essential. The level of "burstiness" of inferred SFHs often varies between authors, even when using CMD data of similar depth. Assumptions about any smoothing that is applied can vary between SFH inversion codes. While this may have less effect on the inferred overall cumulative mass growth over time, it can strongly affect the statistical significance of SFH peaks that may or may not align with orbital events. Also, the exact timing of any burst can vary depending on the assumed stellar models. These uncertainties should all be kept in mind for the comparisons discussed below. Nonetheless, we can still make some general assertions, especially for the galaxies with the most precise orbits and SFHs. This is the first time this type of comparison has been conducted for galaxies at the edge of the LG. Moreover, we can reexamine closer galaxies, such as Leo~I, in more detail than previously possible. 

\subsubsection{First Infall Galaxies}

First-infall galaxies should not show imprints of previous interactions with nearby massive galaxies (i.e., the LMC, M33, MW, or M31 in this work) in their SFHs (i.e., sudden bursts). We do indeed confirm this by comparing the published SFHs \citep[][]{Gallart2015,Albers2019} to those of first infall galaxies from simulations \citep[][]{Garrison-Kimmel2019}. This exercise shows that for most of our galaxies, the SFH matches up well with expectations, as follows.

For the more massive first-infall dwarf galaxies in the sample (M$_{*}$=$1\times 10^{7-9}\, M_{\odot}$, i.e. IC~1613, WLM) star formation has been relatively constant, with the galaxies assembling their stellar mass at a steady rate since early times \citep{Albers2019}. This agrees with the predictions from simulations \citep{Digby_2019,Garrison-Kimmel2019,Joshi_2021}.

For lower mass dwarfs (M$_{*}$=$1\times 10^{5-7}\, M_{\odot}$, i.e. Leo A, Leo T, Pegasus dIrr, Phoenix, Sag DIG, UGC 4879) we see a different pattern with much of the stellar mass being accumulated in the early universe, prior to the epoch of reionization, and low levels of star formation in more recent times \citep[][]{Jacobs_2011,Weisz_2012,Weisz2014,Gallart2015}{}{}. While the overall pattern of SFHs does match between observations and simulations, there are discrepancies in these smaller mass systems. Mainly that the late-time star formation is still larger than expected from simulations, with the level of discrepancy depending on the individual dwarf \citep[][]{Digby_2019,Garrison-Kimmel2019,Joshi_2021}{}{}. 
We cannot shed new light on the cause of this difference in low-mass dwarfs, but can confirm that it is not caused by the presence of nearby massive galaxies. 

The exception to this pattern is Leo A. It has a stellar mass M$_{*}$=$5\times 10^{6}\, M_{\odot}$ \citep[][]{McConnachie_2012}{}{} where we would expect to see most star formation in the early universe with a relatively limited contribution from late times \citep[][]{Digby_2019,Garrison-Kimmel2019}. Instead, Leo A has actually assembled $\sim 85\%$ of its stellar mass in the last 6~Gyrs, with a main maximum centered on 5~Gyrs look-back time \citep[][]{Cole2007,Skillman_2014,Ruiz-Lara_2018}{}{}. 
Such late-time star formation is generally not seen in simulations of dwarf galaxies in a LG environment \citep[][]{Digby_2019,Garrison-Kimmel2019,Joshi_2021,Samuel_2022}, but is seen infrequently in simulations of isolated dwarf galaxies \citep[][]{BL_2015,Joshi_2021}{}{}.

This unusual SFH could be evidence of Leo A being a backsplash galaxy from M31, with the enhancement in star formation being caused by the compression of Leo A's gas reservoir by the hot gas halo of M31. However, we find in this work that such a previous interaction is not consistent with the measured PMs. Therefore, this late-time star formation must be caused by another mechanism, such as a dwarf-dwarf merger \citep[][]{Kirby2017}{}{}, stochastic mass accretion \citep[][]{Ledinauskas2018} or early universe star formation being suppressed by internal mechanisms \citep[][]{Cole2007}{}{}. Our measurements cannot discriminate between these scenarios, but we can rule out that a previous interaction with either the MW or M31 is the cause of the unusual late star formation in Leo A. 

\subsubsection{Interacting \& Backsplash Galaxies}

For each of the backsplash galaxies or galaxies still undergoing an interaction in our sample, we can assess whether SFH features (e.g., quenching or star formation) correspond to specific orbital features (e.g. past virial crossing and pericenters). This will allow us to better understand the influence of the interaction with the host.  

\bigskip

\begin{figure*}[t]
\centering \includegraphics[scale=0.6]{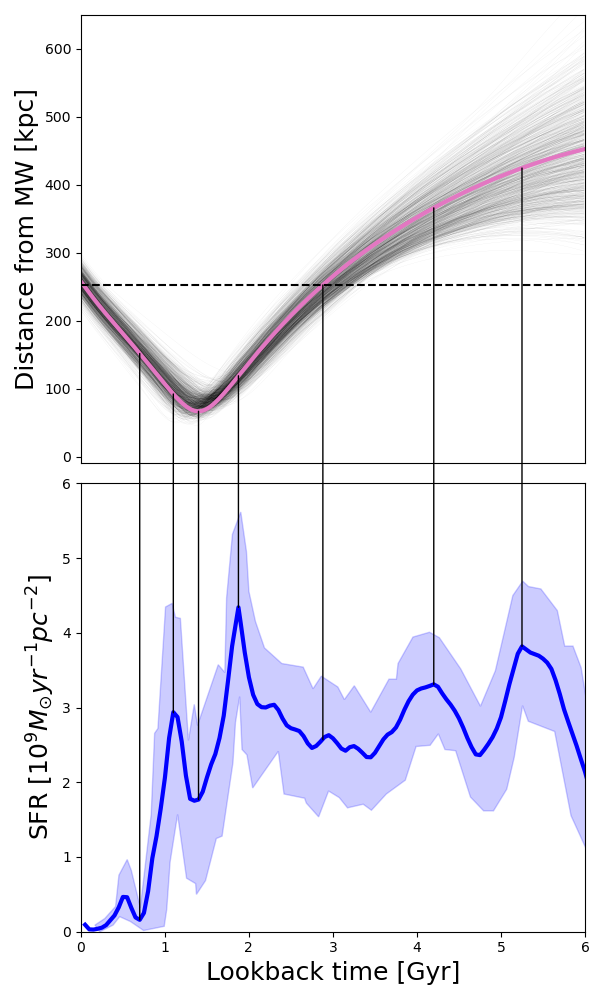}
    \caption{The orbit for Leo~I (Upper panel) compared to its SFH \citep[Lower Panel;][]{Ruiz-Lara_2021}{}{}. The pink line shows the direct orbit for Leo~I and the gray lines show the other orbits drawn from our Monte Carlo ensemble based on the observational uncertainties (see Section \ref{subsec:orbits}). The blue line indicates the star formation rate over time for the past 6 Gyrs with the light blue shape showing the uncertainties. The dashed black line in the upper panel indictaes the virial radius of the MW ($R_{vir} = 251$~kpc) Key times are linked between the panels with black lines.}
    \label{fig:SFH}
\end{figure*}

\noindent \textbf{Leo~I:} This is the backsplash galaxy for which we have the most precise combination of SFH \citep[][]{Ruiz-Lara_2021} and orbital history and is, therefore, the galaxy we can most easily compare to simulations. 
Leo~I has a high rate of star formation in the early universe which is substantially reduced following the epoch of reionization. Though it does not fully quench, this matches the SFH of other dwarfs with similar masses, \citep[M$_{*}$=$5\times 10^{6}\, M_{\odot}$, e.g.~Phoenix;][]{McConnachie_2012,Garrison-Kimmel2019}{}{}.

Figure~\ref{fig:SFH} compares the SFH of Leo~I with the orbital history we derived over the same period of time. The SFH shows a substantial burst of star formation at $\sim$5.3 and $\sim$4.2 Gyr ago. This is long before Leo~I begins to interact with the MW according to our orbital history. Using the metallicity of Leo~I's stars, the  $\sim$5.3 Gyr ago burst has been argued to be caused by a dwarf-dwarf merger where Leo~I accreted a lower mass dwarf galaxy \citep[][]{Ruiz-Lara_2021}. These intense bursts of star formation continued until $\sim$3 Gyrs ago when Leo~I crossed the MW's virial radius. This is followed by a steady rise in SFR until a peak $\sim$2 Gyr ago. This rise is likely caused by the compression of gas in Leo~I producing a starburst before the gas is stripped by the MW. 
Following this, the SFR falls rapidly as Leo~I approaches the pericenter of its orbit ($\sim$1.7 Gyrs ago). It then experiences another smaller burst as it begins to move away from the MW ($\sim$1.3 Gyrs ago). Following this burst the SFR falls until Leo~I fully quenches relatively recently ($\sim$300 Myrs ago) as it is leaving the MW virial radius. 

This strongly suggests that the MW interaction has had a significant impact on Leo~I and led to a starburst followed by quenching. However, the timescale of this interaction is significantly long with Leo~I remaining star forming for many Gyrs after falling into the virial radius of the MW. 
This is important to note for comparison both to simulations and to galaxies outside the LG. 


This detailed timeline, linking the SFH and position of Leo~I throughout its interaction with the MW can be compared to simulations and give new constraints on future modeling. 
If we compare to the FIRE simulations we find that the SFH for Leo~I-like dwarfs are stripped quickly ($<$2~Gyrs) after falling into a MW mass host \citep{Garrison-Kimmel2019,Samuel_2022}. Whereas in our comparison we find that Leo~I remained star forming for almost 3~Gyrs after falling into the MW's virial radius. This indicates that gas stripping may be too efficient in these simulations compared to observations \citep[][]{Emerick2016}{}{}. This finding would also match with observations of the satellite systems of MW analog galaxies in the Local Volume \citep{Mao_2021,Carlsten_2022,Karunakaran_2023,Jones2023} which show that Leo~I mass galaxies are often ($>$60\%) not quenched even within the virial radius of their hosts. 

\bigskip

\noindent \textbf{Leo~II:} This galaxy was already inside the MW virial radius at the point our orbital history starts, 6~Gyrs ago. Our orbit shows it has been around the edge of the MW virial radius ever since. Its SFH shows that is has been fully quenched with no additional star formation for the entire 6~Gyrs \citep[][]{Weisz2014}{}{}. Hence, we cannot assess whether Leo~II's SFH quenching coincides with its first entry into the MW virial radius, since this happened earlier than our orbits can reliably trace.

\bigskip

\noindent \textbf{NGC~6822:} This galaxy is a special case. While it has potentially interacted with the MW, as discussed above, in most orbits consistent with our measurements this interaction is brief and limited to the edge of the MW's virial radius (see Figure~\ref{fig:errors}). Therefore we expect any influence from the MW on the SFH of NGC~6822 to be limited to a relatively small time window. The SFH of NGC 6822 was studied by \citet[][]{Weisz2014} and \citet[][]{Fusco2014}. The former authors find that the star formation started to increase significantly $\sim 6$ Gyr ago, a time at which our orbits calculation indicate the galaxy was still far from the MW. After that time, the \citet[][]{Weisz2014} SFH shows some indication for individual bursts at both $\sim 4$ and $\sim 2$ Gyr ago. \citet[][]{Fusco2014}, for their field closest to the galaxy center, find a general increase in the star formation rate between $5$ and $1.5$ Gyr ago, but not necessarily in the form of individual bursts. Either way, both of these studies find an increase in star formation around the time when NGC 6822 may have passed closest to the MW (see Figure~\ref{fig:errors}). So perhaps this is indicative of a potential interaction between NGC~6822 and the MW, although deeper CMDs that reach the oMSTO would be desirable to further substantiate this. 

\bigskip

\noindent \textbf{IC~10:} The SFH of IC~10 was studied by \citet[][]{Weisz2014}. \citet[][]{DellAgli2018} then used this SFH as a basis, and refined it using the infrared properties of evolved stars as an additional constraint. 
These combined studies imply that the major epoch of star formation that occurred after reionization was $\sim 2.5$ Gyr ago. 
Then there was additional star formation between $\sim 0.3$--$1$ Gyr ago, as well as ongoing star formation over the past $\sim 40$ Myr. In our orbit calculations (see Figure~\ref{fig:orbits}), IC~10 crosses the M31 virial radius $\sim$3 Gyrs ago. This is just prior to the inferred peak of star formation, which was likely caused by the compression of gas. The star formation then decreases as IC~10 approaches pericenter $\sim$ 1.3~Gyrs ago, but there are then additional bursts as the galaxy moves away from its pericenter. IC~10 is almost 10$\times$ more massive than Leo~I \citep[][]{McConnachie_2012}{}{} and therefore this may explain why it has better retained its gas reservoir throughout this interaction. This could be why IC~10 is a starburst galaxy, while Leo~I has quenched, despite IC~10 having longer exposure to the halo of M31 compared to Leo~I and the MW. 
While our findings for IC~10 are broadly similar to the case of Leo~I, and based exclusively on relatively recent star formation, deeper CMDs that reach the oMSTO would again be highly desirable to further substantiate these findings. 
 
IC~10 is not a true backsplash galaxy at the present time, since it still is inside the virial radius of M31 ($R_{\rm vir}$ = 329 kpc). However, our orbits predict it will move outside the virial radius $\sim$1 Gyr in the future (see Figure \ref{fig:orbits}). Therefore IC~10 is a likely to become a backsplash galaxy in the near-future.

\bigskip

\subsection{Future Improvements} \label{subsec:future}

Various improvements on this work may be possible in the future. 

\smallskip

\noindent {\it Galaxy sample size:} There do exist other distant dwarf galaxies with suitable first-epoch \textit{HST} data for which a PM determination by comparison to \textit{Gaia} may in principle be possible. For example, a galaxy like NGC~205 does show an overdensity in the relative PM frame. However, there are insufficient \textit{Gaia} stars that can act as an anchor for the absolute reference frame. This might improve either with future \textit{Gaia} data releases, or with more sophisticated treatments of frame alignments between \textit{HST} and \textit{Gaia} than those performed by \gaiahub{} \citep[e.g.][]{McKinnon_2023}{}{}.   

\smallskip

\noindent {\it PM accuracy:} 
This will improve for \gaiahub{} measurements from future \textit{Gaia} data releases, both because of the increased \textit{Gaia} positional accuracy, and because of the increased time baseline relative to the existing \textit{HST} data. Moreover, for 2 additional dwarf galaxies in our sample, Peg dIrr and Leo T, we expect to produce multi-epoch \textit{HST} PMs based on a second epoch from \textit{HST}-GO-17174 in a similar manner to Leo A and Sag DIG in this work. These particular results will be a substantial improvement over the PMs derived from \gaiahub{} in this work and provide more detail on the orbital history of these first infall galaxies. 


\smallskip

\noindent {\it Data-Model comparisons:} Future projects, such as that approved in \textit{HST}-AR-16628 (PI: E. Patel), aim to develop quantitative connections between orbital histories and SFHs for those galaxies that have precise measurements/predictions for both. In particular, it is necessary to develop tools that move beyond the typical two-point correlation used to connect quenching and infall time. Instead, this program aims to move towards using these rich data sets to paint a clear picture of where, when, and how low-mass MW satellite galaxies quench by building a quantitative framework that can link together orbits and SFHs across the last few billion years.

\section{Summary} \label{sec:Con}

We have determined the PMs of 12 dwarf galaxies in the LG, ranging from the outer MW halo to the edge of the LG, using the \gaiahub{} software that measures PMs using \textit{HST} as the first and \textit{Gaia} as the second epoch. These PMs provide an improvement in precision over previous work using \textit{Gaia} DR3 for 9 of the 12 galaxies. We have also determined even more precise (5-8x smaller errors) PMs for two of the galaxies in our sample, Leo A and Sag DIG, using multi-epoch \textit{HST} imaging, based on comparing the movement of member stars relative to background galaxies. 

We have used these PMs, combined with literature distances, and line-of-sight velocities, to derive orbital histories for these dwarfs. These calculations take into account the gravitational potential of all the massive galaxies of the LG (the MW, M31, M33 and the LMC). The orbital histories show that most of the galaxies in our sample, 8 of 12, are on first infall into their LG host galaxy with $>$90\% certainty (see Table \ref{tab:splash}). 

We then compare the SFHs of these first infall dwarfs to those from simulations. This shows that the dwarfs are consistent with simulated dwarfs across most of the sample. High-mass (M$_{*}$=$1\times 10^{7-9}\, M_{\odot}$) infalling dwarfs show approximately constant star formation across their lifetimes. Low-mass infalling dwarfs (M$_{*}$=$1\times 10^{5-7}\, M_{\odot}$) instead show most star formation in the early universe, before reionization, with more limited star formation in late times. The one exception to this pattern is Leo A. This is a low-mass galaxy that formed the majority of its stars in the past $\sim$5~Gyrs. While this unusual star formation pattern could in principle be due to a previous interaction with M31, the orbit implied by our new PMs does not support this.  

The remaining 4 galaxies in our sample that are not on first infall  have had interactions with either the MW (Leo~I, Leo~II and NGC~6822), or M31 (IC~10). When we compare their star formation and orbital histories we find general agreement with the following scenario (but with the strength of the conclusions limited by the observational uncertainties in both the orbital and star formation histories). Star formation is triggered after the dwarf crosses the virial radius of its host. This star formation then peaks and decreases as the dwarf nears its pericenter. After the dwarf moves away from its pericenter, there is continued quenching, interspersed by additional smaller bursts. Overall, the quenching process is not fast, and can take several Gyrs. This scenario is best exemplified by the case of Leo~I, for which both the orbital and star formation histories are tightly constrained.

These results demonstrate that even relatively small dwarfs like Leo~I can hold on to HI reservoirs and avoid quenching for several Gyrs after falling into the virial radius of their host. This is longer than is generally found in simulations, for which Leo~I-mass dwarfs quench rapidly after falling into the influence of a massive galaxy. A longer quenching timescale is consistent with recent findings around MW-analogs within the Local Volume. These show that Leo~I mass dwarfs are only quenched $\sim$40\% of the time even when projected within the virial radius of their host galaxy. 


\begin{acknowledgments}

Support for this work was provided by grants for \textit{HST} archival programs AR-15633 and 16628, and GO program 15911 and 17174 provided by the Space Telescope Science Institute (STScI). EP acknowledges financial support provided by NASA through the Hubble Fellowship grant \# HST-HF2-51540.001-A awarded by STScI. STScI is operated by the Association of Universities for Research in Astronomy, Incorporated, under NASA contract NAS5-26555.

A. del Pino acknowledges the financial support from the European Union - NextGenerationEU and the Spanish Ministry of Science and Innovation through the Recovery and Resilience Facility project J-CAVA and the project PID2021-124918NB-C41.

A. Aparicio acknowledges financial support from Ministry of Science and Innovation of Spain through the grant PID2020-115981GB-I00

C.G.  acknowledges support from the Agencia Estatal de Investigación del Ministerio de Ciencia e Innovación (AEI-MCINN) under grant “At the forefront of Galactic Archaeology: evolution of the luminous and dark matter components of the Milky Way and Local Group dwarf galaxies in the Gaia era” with reference PID2020-118778GB-I00/10.13039/501100011033 and from the Severo Ochoa program through CEX2019-000920-S.

This work has made use of data from the European Space Agency (ESA) mission {\it Gaia} (\url{https://www.cosmos.esa.int/gaia}), processed by the {\it Gaia} Data Processing and Analysis Consortium (DPAC, \url{https://www.cosmos.esa.int/web/gaia/dpac/consortium}). Funding for the DPAC has been provided by national institutions, in particular the institutions participating in the {\it Gaia} Multilateral Agreement.

This research made use of Astropy,\footnote{https://www.astropy.org} a community-developed core Python package for Astronomy \citep{astropy2013, astropy2018}.

This research has made use of NASA's Astrophysics Data System Bibliographic Services.

This project is part of the HSTPROMO (High-resolution Space Telescope PROper MOtion) Collaboration\footnote{\url{http://www.stsci.edu/~marel/hstpromo.html}}, a set of projects aimed at improving our dynamical understanding of stars, clusters, and galaxies in the nearby Universe through measurement and interpretation of proper motions from \textit{HST}, \textit{Gaia}, and other space observatories. We thank the collaboration members for sharing their ideas and software.

\end{acknowledgments}

\bibliography{refs}{}
\bibliographystyle{aasjournal}

\end{document}